\documentclass[useAMS,usenatbib]{mn2e}
\usepackage{color}
\usepackage{colortbl}
\usepackage{dcolumn}
\usepackage{amsmath}
\usepackage{amssymb}
\usepackage{graphicx}
\usepackage{epsfig}
\usepackage{changebar}
\usepackage{natbib}
\usepackage{multirow}
\usepackage{subfigure}
\usepackage{txfonts}
\usepackage{relsize}
\usepackage{verbatim} 
\usepackage{natbib}
\usepackage{rotating}
\usepackage{lscape}
\usepackage{graphicx}
\newcommand{\kms}{km\,s$^{-1}$}

\newcommand{\aap}{A\&A}

\newcommand{\aaps}{A\&AS}
\newcommand{\mnras}{MNRAS}
\newcommand{\apj}{ApJ}
\newcommand{\apjs}{ApJS}
\newcommand{\apjl}{ApJL}

\newcommand{\pasa}{PASA}
\newcommand{\aj}{AJ}

\newcommand{\ssr}{Science Reviews}
\newcommand{\memras}{Royal Astronomical Society, Memoirs}
\newcommand{\aapr}{A\&AR}

\newcommand{\memsai}{MmSAI}
\newcommand{\affil}[1]{$^{\rm #1}$}
\newcommand{\physrep}{Phys. Rep.}
\newcommand{\na}{New Astronomy}
\newcounter{inst} 
\newcommand{\inst}[1]{\noindent%
   \refstepcounter{inst}\affil{\arabic{inst}\label{#1}}     
   }

\title[MWA observations of A3667]{The first Murchison Widefield Array low frequency radio observations of cluster scale non-thermal emission: the case of Abell 3667}
\author[L.Hindson]{\normalsize L. Hindson$^*$\affil{\ref{VUW}}, M. Johnston-Hollitt\affil{\ref{VUW}}, N. Hurley-Walker\affil{\ref{ICRAR}}, K. Buckley\affil{\ref{VUWCOMP}}, J. Morgan\affil{\ref{ICRAR}}, E.Carretti\affil{\ref{CSIRO}}, K.S. Dwarakanath\affil{\ref{RRI}}, M. Bell\affil{\ref{CSIRO},\ref{CAASTRO},\ref{USyd}}, 
G.~Bernardi\affil{\ref{CfA},\ref{SKA},\ref{SKA2}}, \newauthor \normalsize
R.~Bhat\affil{\ref{ICRAR},\ref{CAASTRO}}, 
J.~D.~Bowman\affil{\ref{ASU}},
F.~Briggs\affil{\ref{CAASTRO},\ref{ANU}}, 
R.~J.~Cappallo\affil{\ref{Haystack}},
B.~E.~Corey\affil{\ref{Haystack}},
A.~A.~Deshpande\affil{\ref{RRI}}, 
D.~Emrich\affil{\ref{ICRAR}},
A.~Ewall-Wice\affil{\ref{MIT}},
L.~Feng\affil{\ref{MIT}}, 
B.~M.~Gaensler\affil{\ref{CAASTRO},\ref{USyd}},\newauthor \normalsize
R.~Goeke\affil{\ref{Haystack}},
L.~J.~Greenhill\affil{\ref{CfA}},
B.~J.~Hazelton\affil{\ref{UDub}}, 
D.~Jacobs\affil{\ref{ASU}},
D.~L.~Kaplan\affil{\ref{UWM}},
J.~C.~Kasper\affil{\ref{CfA}},
E.~Kratzenberg\affil{\ref{Haystack}},
N. Kudryavtseva\affil{\ref{ICRAR}}, 
E.~Lenc\affil{\ref{CAASTRO},\ref{USyd}},\newauthor \normalsize
C.~J.~Lonsdale\affil{\ref{Haystack}},
M.~J.~Lynch\affil{\ref{ICRAR}},
S.~R.~McWhirter\affil{\ref{Haystack}},
B.~McKinley\affil{\ref{CAASTRO},\ref{ANU}}, 
D.~A.~Mitchell\affil{\ref{CASS},\ref{CAASTRO}},
M.~F.~Morales\affil{\ref{UDub}},
E.~Morgan\affil{\ref{MIT}},
D.~Oberoi\affil{\ref{NCRA}},
S.~M.~Ord\affil{\ref{ICRAR},\ref{CAASTRO}},\newauthor \normalsize
B.~Pindor\affil{\ref{CAASTRO},\ref{UMelb}}, 
T.~Prabu\affil{\ref{RRI}},
P.~Procopio\affil{\ref{CAASTRO},\ref{UMelb}}, 
A.~R.~Offringa\affil{\ref{CAASTRO},\ref{ANU}}, 
J.~Riding\affil{\ref{CAASTRO},\ref{UMelb}},
A.~E.~E.~Rogers\affil{\ref{Haystack}},
A.~Roshi\affil{\ref{NRAO}},
N.~Udaya~Shankar\affil{\ref{RRI}},
K.~S.~Srivani\affil{\ref{RRI}},\newauthor \normalsize
R.~Subrahmanyan\affil{\ref{RRI},\ref{CAASTRO}},
S.~J.~Tingay\affil{\ref{ICRAR},\ref{CAASTRO}},
M.~Waterson\affil{\ref{ICRAR},\ref{ANU}},
R.~B.~Wayth\affil{\ref{ICRAR},\ref{CAASTRO}},
R.~L.~Webster\affil{\ref{CAASTRO},\ref{UMelb}},
A.~R.~Whitney\affil{\ref{Haystack}},
A.~Williams\affil{\ref{ICRAR}},
C.~L.~Williams\affil{\ref{MIT}}
 \\
\\%
{\smaller \affil{}\,$^*$Email: Luke.Hindson@vuw.ac.nz}\\
{\smaller \inst{VUW}\,School of Chemical and Physical Sciences, Victoria University of Wellington, Wellington, New Zealand}\\
{\smaller \inst{ICRAR}\,ICRAR - Curtin University, Perth, Australia}\\
{\smaller \inst{VUWCOMP}\,School of Engineering and Computer Sciences, Victoria University of Wellington, Wellington, New Zealand}\\
{\smaller \inst{CSIRO}\,CSIRO Astronomy and Space Science, PO Box 76, Epping, NSW 1710, Australia}\\
{\smaller \inst{RRI}\,Raman Research Institute, Bangalore, India}\\
{\smaller \inst{ASU}\,School of Earth and Space Exploration, Arizona State University, Tempe, AZ, USA}\\
{\smaller \inst{NCRA}\,National Centre for Radio Astrophysics, Pune, India}\\
{\smaller \inst{CAASTRO}\,ARC Centre of Excellence for All-sky Astrophysics (CAASTRO), Redfern, NSW, Australia}\\
{\smaller \inst{USyd}\,Sydney Institute for Astronomy, School of Physics, The University of Sydney, NSW 2006, Australia}\\
{\smaller \inst{CfA}\,Harvard-Smithsonian Center for Astrophysics, Cambridge, MA, USA}\\
{\smaller \inst{ANU}\,Research School of Astronomy and Astrophysics, The Australian National University, Canberra, Australia}\\
{\smaller \inst{MIT}\,MIT Kavli Institute for Astrophysics and Space Research, Cambridge, MA, USA}\\
{\smaller \inst{UDub}\,Physics Department, University of Washington, Seattle, WA, USA}\\
{\smaller \inst{UWM}\,Physics Department, University of Wisconsin-Milwaukee, Milwaukee, WI, USA}\\
{\smaller \inst{CASS}\,CSIRO Astronomy and Space Science, PO Box 76, Epping, NSW 1710, Australia}\\
{\smaller \inst{UMelb}\,School of Physics, The University of Melbourne, Melbourne, Australia}\\
{\smaller \inst{Haystack}\,MIT Haystack Observatory, Westford, MA, USA}\\
{\smaller \inst{NRAO}\,National Radio Astronomy Observatory, Charlottesville, WV, USA}\\
{\smaller \inst{SKA}\,Square Kilometre Array South Africa, Pinelands, Cape Town, South Africa}\\
{\smaller \inst{SKA2}\,Department of Physics and Electronics, Rhodes University, Grahamstown, South Africa}\\
}
\newpage

\begin{document}

\pagerange{\pageref{firstpage}--\pageref{lastpage}} \pubyear{2010}

\maketitle

\label{firstpage}

\begin{abstract}
We present the first Murchison Widefield Array observations of the well-known cluster of galaxies Abell\,3667 (A3667) between 105 and 241\,MHz. A3667 is one of the best known examples of a galaxy cluster hosting a double radio relic and has been reported to contain a faint radio halo and bridge. The origins of radio halos, relics and bridges is still unclear, however galaxy cluster mergers seems to be an important factor. We clearly detect the North-West (NW) and South-East (SE) radio relics in A3667 and find an integrated flux density at 149\,MHz of $28.1\pm1.7$ and $2.4\pm0.1$\,Jy, respectively, with an average spectral index, between 120 and 1400\,MHz, of $-0.9\pm0.1$ for both relics. We find evidence of a spatial variation in the spectral index across the NW relic steepening towards the centre of the cluster, which indicates an ageing electron population. These properties are consistent with higher frequency observations. We detect emission that could be associated with a radio halo and bridge. However, due to the presence of poorly sampled large-scale Galactic emission and blended point sources we are unable to verify the exact nature of these features. 

\end{abstract}


\section{Introduction}

Observations of galaxy clusters in the radio regime have revealed low-surface brightness and large-scale non-thermal emission up to Mpc-scales associated with the periphery and central regions of some galaxy clusters (see, e.g., \citealt{Ferrari2008,Cassano2009,Venturi2011, Feretti2012, Brunetti2014} for reviews). The origin of this radio emission does not appear to be directly associated with the activity of individual galaxies within the cluster but rather pervades the intracluster medium (ICM). These radio sources are known as radio halos and relics and share some observational properties: they both have steep synchrotron spectra\footnote{where $S_\nu \propto \nu^\alpha$ and $\alpha$ is the spectral index.} ($\alpha$ $\sim -1.0$ to $-1.4$) and are characterised by very low-surface brightness emission ($\sim\umu$\,Jy\,arcsec$^{-2}$ at 1.4\,GHz). Radio halos and relics differ in their polarisation properties and locations within galaxy clusters: radio halos are unpolarised, at current detection levels \citep{Govoni2013}, and located towards the centre of the host galaxy cluster, whilst relics are highly polarised and reside on the periphery of the galaxy cluster. The presence of radio halos and relics within galaxy clusters is direct evidence that magnetic fields and relativistic particles exist within the ICM. 

Radio relics are believed to be generated by the acceleration of electrons by the diffusive shock acceleration mechanism (first-order Fermi process) in shocks propagating to the cluster periphery after a major merger event \citep{Ensslin1998}. These electrons are embedded within the cluster wide magnetic fields and so emit synchrotron radiation. 

The mechanism by which radio halos emit and maintain the observed synchrotron emission is a topic of debate. This is because electrons emitting via the synchrotron mechanism have limited lifetimes and some as yet unknown mechanism must exist to (re-)accelerate the electrons within the ICM to maintain the observed levels of synchrotron emission in radio halos. Any theory that seeks to explain the origin of radio halos within the ICM must be able to describe how relativistic electrons are reaccelerated. Two mechanisms are favoured in the literature: primary acceleration of electrons due to turbulence \citep{Jaffe1977} and secondary production of electron/positron pairs due to hadronic collisions between cosmic-ray protons and thermal protons \citep{Dennison1980}. Sensitive low-frequency observations of a large sample of clusters should be able to distinguish which of these models is responsible for generating radio halos  \citep{Cassano2010B,Cassano2012}. One of the primary observable differences between these two models is that the turbulent re-acceleration model predicts a population of radio halos with ultra-steep spectra ($\alpha <-1.5$). To date radio halos have been commonly observed at $\sim$\,GHz frequencies where emission from ultra-steep spectrum halos is weak and easily missed. Observations at low frequency should be able to detect these ultra-steep spectrum radio halos. 

In a small number of clusters a ``bridge'' of radio emission has been observed connecting relics on the periphery of the cluster with the halo at the centre. These radio bridges are characterised by a filament of low surface brightness emission that extends over hundreds of kilo-parsecs. Only a few examples of radio bridges have been observed such as in the Coma cluster, A2255, A2744 \citep{Giovannini2004} and recently in A3667 \citep{Carretti2013}. The origin of radio bridges is unknown; they could simply be the result of projection effects whereby extended radio emission is seen projected against the galaxy cluster. Alternatively, a radio bridge may be a feature that has formed alongside the radio relic during a merger event and is the result of shocks and turbulence accelerating the local electron population \citep{Markevitch2010}. 

To date it has been difficult to construct a large sample of these radio features due to a lack of instrumentation capable of surveying a large extent of the sky at low frequency with adequate sensitivity. With the advent of instruments such as the Murchison Widefield Array (MWA; \citealt{Tingay2013,Bowman2013}) and the LOw Frequency ARray (LOFAR; \citealt{Haarlem2013}) we now have the ability to carry out such observations. In this study we aim to observe a single galaxy cluster that is reported to host all three of these extended radio features, at low frequency using the MWA, as a preliminary step towards further such detections.

The target of our study is the well-known galaxy cluster A3667 ($\alpha=302\fdg86$, $\delta=-56\fdg72$). Optical studies of the galaxy members within A3667 \citep{Proust1988,Sodre1992,Girardi1996,Girardi1998,JohnstonHollitt2008, Owers2009} have revealed that A3667 is relatively nearby with an average redshift of $z=0.0553$ \citep{Owers2009}. Optical observations have also revealed a high galaxy density within A3667, with an elongated spatial distribution of galaxies in the SE to NW direction. Many authors propose that this elongation coupled with the high velocity dispersion of cluster members ($\sim 1200$\,\kms) is an indication of a recent or ongoing merger between two sub-clusters in the plane of the sky. Numerous X-ray observations of A3667 \citep{Knopp1996,Ebeling1996,Markevitch1999, Vikhlinin2001,Finoguenov2010,Sarazin2013} have probed the physical properties of the hot gas that exists within the ICM of A3667. The cluster is found to have an X-ray luminosity of $L_{\rm x}=9\times10^{44}$\,ergs\,s$^{-1}$ in the 0.1--2.4\,keV range \citep{Ebeling1996} with a temperature profile that is also elongated in the SE to NW direction and shows a sharp change in temperature at the NW periphery of the cluster. These X-ray features also suggest that A3667 has undergone a merger event and that shocked regions should be present on the periphery of the cluster \citep{Vikhlinin2001,Markevitch1999}. 

The most prominent features of A3667 are two large and bright radio relics located on the NW and SE periphery of the cluster. These NW and SE relics have integrated flux densities of $3.70\pm0.30$ and $0.30\pm0.02$\,Jy at 1.4\,GHz, respectively, \citep{JohnstonHollitt2003} and extend over approximately $0\fdg5$ and $0\fdg2$, respectively. Numerous radio observations of A3667 (e.g., \citealt{Mills1961,Ekers1969,Schilizzi1975,Goss1982,Rottgering1997}) have noted the bright extended structures in A3667 but failed to image the entire extent of both relics. The first study of the entire region at high resolution ($\sim6\arcsec$), including both the NW and SE relics, was carried out by \citet{JohnstonHollitt2003} at 1.4\,GHz. These observations revealed a complex filamentary flux density distribution with an average spectral index, between 0.8 and 1.4\,GHz, of $-0.9\pm0.2$ and $-1.2\pm0.2$ for the NW and SE relics, respectively. The spectral index was found to vary considerably over both relics with a range of $-0.5$ to $-1.4$ for the NW relic and $-0.5$ to $-2.1$ for the SE relic. The spectral index of the SE relic is found to have a spatial gradient with a flatter spectral index at the leading edge of the relic, steepening towards the centre of the cluster \citep{JohnstonHollitt2003}. Recent observations of A3667 at 2.3 and 3.3\,GHz report the detection of a putative radio bridge connecting the NW relic to the centre of the cluster \citep{Carretti2013}. The presence of a faint radio halo in the centre of the cluster has also been reported \citep{JohnstonHollitt2004,Carretti2013}. The only study of A3667 below 300\,MHz was performed more than five decades ago by \cite{Mills1961} at 86\,MHz using the Mills Cross Telescope with a resolution of $50\arcmin$. The large angular extent and high surface brightness of the relics and the detection of a faint radio halo and bridge (at frequencies of 1.4--3.3\,GHz) in A3667 make it an excellent laboratory in which to carry out low frequency observations to study the ICM.

In this paper we present the first MWA observations of A3667 between 105 and 241\,MHz. The layout of this paper is as follows: in Section~\ref{Sect:obs} we present the observations and discuss the data reduction process. In Section~\ref{Sect:results} we present the results of these observations including images of the radio data, integrated flux density estimates of emission features and a spectral index map of A3667. In Section~\ref{Sect:discussion} we discuss our results for the NW and SE radio relics and search for the presence of a radio halo and bridge in A3667. We end with a brief outline of future prospects with the MWA and our conclusions in Section~\ref{Sect:summary}. Throughout this paper we assume the following cosmological parameters: $H_{\rm 0}=73$\,km\,s$^{-1}$\,Mpc$^{-1}$, $\Omega_{\rm m}=0.27$ and $\Omega_\Lambda = 0.73$. This implies that at the redshift of A3667 (0.0553) $1\arcmin= 61.86$\,kpc.

\section{Observations \& data reduction procedures}\label{Sect:obs}

The MWA is a new radio telescope operating in the low-frequency regime between 80 and 300\,MHz. Details of the technical design and specifications of the MWA is given by \cite{Tingay2013} and a review of the performance properties and scientific goals is presented by \cite{Bowman2013}. Consisting of 128 ``tiles'' of $4\times4$ dipole antennas, which are concentrated in a dense core $<1.5$\,km in diameter in order to achieve high surface brightness sensitivity.

\begin{figure}
\includegraphics[width=0.5\textwidth]{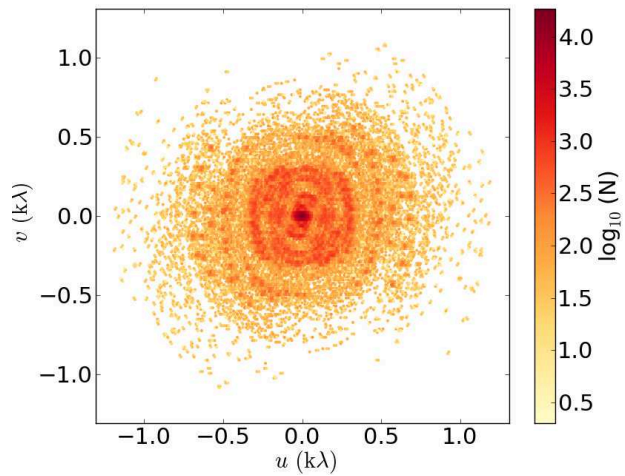} 
\caption{Hexagonal binning plot showing the density of $u,v$ points for a combined $3\times232$\,s observation centred on a single channel at 149\,MHz. Each hexagonal bin has a width of 0.006\,k$\lambda$ and the colour scale indicates the logarithm of the number of $u,v$ points within each hexagonal bin.}
\label{im:uvcoverage}
\end{figure}

A3667 was observed with the MWA on the 9$^{\rm th}$ of June 2013 using 114 tiles of the full 128-tile array. At the time of observation, these observations utilised the most complete array to date and represent a good approximation of the capabilities of the complete MWA. The observations comprise of four 30.720\,MHz wide sub-bands centred on 120.335, 149.135, 179.855 and 225.935\,MHz (hereafter 120, 149, 180 and 226\,MHz). The MWA forms beams on the sky using analogue beam formers \citep{Tingay2013} whereby the beam is moved by changing the beam former delays by incremental amounts. This moves the primary beam centre and results in a change in the overall shape of the beam response. Each frequency band was observed consecutively with the same delay settings before repeating with new delays to keep A3667 in the centre of the field. Snapshot observations of $3\times232$ seconds were obtained at each central frequency between an hour angle of 21 and 22 hours. The minimum and maximum baselines were 7.7 and 2864\,m, respectively, which provides sensitivity to structures from $\sim 600\arcmin$ to $\sim3\arcmin$. The combined $u,v$-coverage of three snapshots at 149\,MHz, excluding multi-frequency synthesis (MFS), can be seen in Fig.~\ref{im:uvcoverage}. This figure uses hexagonal binning to visualise the $u,v$ plane which contains $>2\times10^6$ $u,v$ points. Hexagonal binning tessellates the $u,v$ plane using a regular grid of hexagons, the number of $u,v$ points that fall within each hexagonal bin is counted and a logarithmic colour-scale is used to represent the total number of $u,v$ points within each hexagonal bin. The extensive $u,v$-coverage at short baselines provides excellent sensitivity to large-scale structure, however the maximum baseline limits the resolution of these observations to $\sim3\arcmin$. The field of view (FoV) of the primary beam at the full width half maximum (FWHM) is approximately $16\degr$ and $10\degr$ at 120 and 226\,MHz, respectively.

\begin{table}
\caption[Observation characteristics]{Summary of the image properties of the combined $3\times232$\,s observations.}
\begin{center}
  \begin{tabular}{cccc}
   \hline Centre Frequency & FWHM of FoV & Resolution & Sensitivity\\
           (MHz) & degrees ($\degr$) & arc minutes ($\arcmin$) & (mJy\,beam$^{-1}$) \\
 	\hline
120	&	15.9	&	$5.3\times3.8$	&	83.3	\\
149	&	13.2	&	$4.2\times3.0$	&	50.0	\\
180	&	11.2	&	$3.6\times2.5$	&	39.5	\\
226	&	9.8		&	$3.0\times2.1$	&	35.1	\\
\hline
\end{tabular}
\end{center}
\label{tab:obs}
\end{table}

The calibration, imaging and deconvolution of these data was performed following the approach of \cite{Hurley2014}. The raw data from the correlator were first processed through {\sc cotter}, a pre-processing pipeline designed to perform several actions. This includes: initial flagging and RFI excision using {\sc aoflagger} \citep{Offringa2010,Offringa2012}; data-averaging to 1\,s time and 40\,kHz frequency resolution and conversion of the data into the measurement set format readable by {\sc casa\footnote{http://casa.nrao.edu/}} (Common Astronomy Software Applications). A single complex gain solution was determined for each tile and frequency using the {\sc bandpass} routine available in {\sc casa} on a single pointed observation of 3C444 taken at the beginning of the A3667 observations. Initial imaging revealed that Centaurus\,A and 3C444, which are approximately 100 and 45 degrees away from the field centre, respectively, were in the primary beam sidelobes of the 180 and 226\,MHz band images. These sources are bright enough to contaminate the primary FoV with stripes caused by sidelobes of the synthesised beam. These sources were peeled from the visibilities by phasing to the sources, determining independent gain solutions and subtracting models of the sources from the $u,v$ data. 

Transforming from the $u,v$ to image plane was performed by converting the calibrated data to a {\sc miriad} \citep{Sault1996} readable format and using {\sc miriad's} {\sc invert} task. Two linear polarisation images (Stokes XX and YY) were produced for each snapshot with a pixel size of $0.5\arcmin$ and an image area of 4800\,pixels$^2$. A ``Briggs'' robust weighting \citep{Briggs1995} of 0.0 was used to give a good compromise between the high sensitivity provided by natural weighting while almost maintaining the synthesised beam of uniform weighting. This weighting scheme allows us to effectively sample the extended emission associated with A3667 without interference from extended emission on larger scales associated with the Galactic plane. The Stokes XX and YY images for each snapshot were deconvolved using the Cotton-Schwab clean algorithm \citep{Schwab1984} using the {\sc miriad} task {\sc clean} to a cut-off level of 3$\sigma$, where $\sigma$ is the root mean squared (RMS) noise. Each XX and YY snapshot was then weighted by the corresponding primary beam and co-added to give the final Stokes I images. {\sc miriad} was chosen to perform imaging and deconvolution because it was found to be considerably faster than the corresponding algorithm in CASA ({\sc clean}). The wide FoV of the MWA means that the $w$-term in the standard two-dimensional Fourier transform equation can no longer be assumed to be zero. The varying $w$-term of each snapshot leads to a systematic positional offset increasing with distance from the centre of the field. To account for this effect the coordinate system of the images was warped using slant orthographic projection \citep{Perley1999,Calabretta2002}. The flux density scale in each snapshot was determined by using an interpolation method described in detail in Appendix\,A before stacking each snapshot together to obtain the final Stokes I images. After performing flux density scaling we find a residual percentage flux density error of 9.1\%, 9.4\%, 4.9\% and 10.8\% at 120, 149, 180 and 226\,MHz, respectively. The final Stokes I images have noise levels of 83.3, 50.0, 39.5 and 35.1\,mJy\,beam$^{-1}$ at 120, 149 and 180 and 226\,MHz, respectively. The image properties are summarised in Table~\ref{tab:obs}. 

\section{Results \& Analysis}\label{Sect:results}

\subsection{Radio images}

\begin{figure*}
\includegraphics[width=0.95\textwidth]{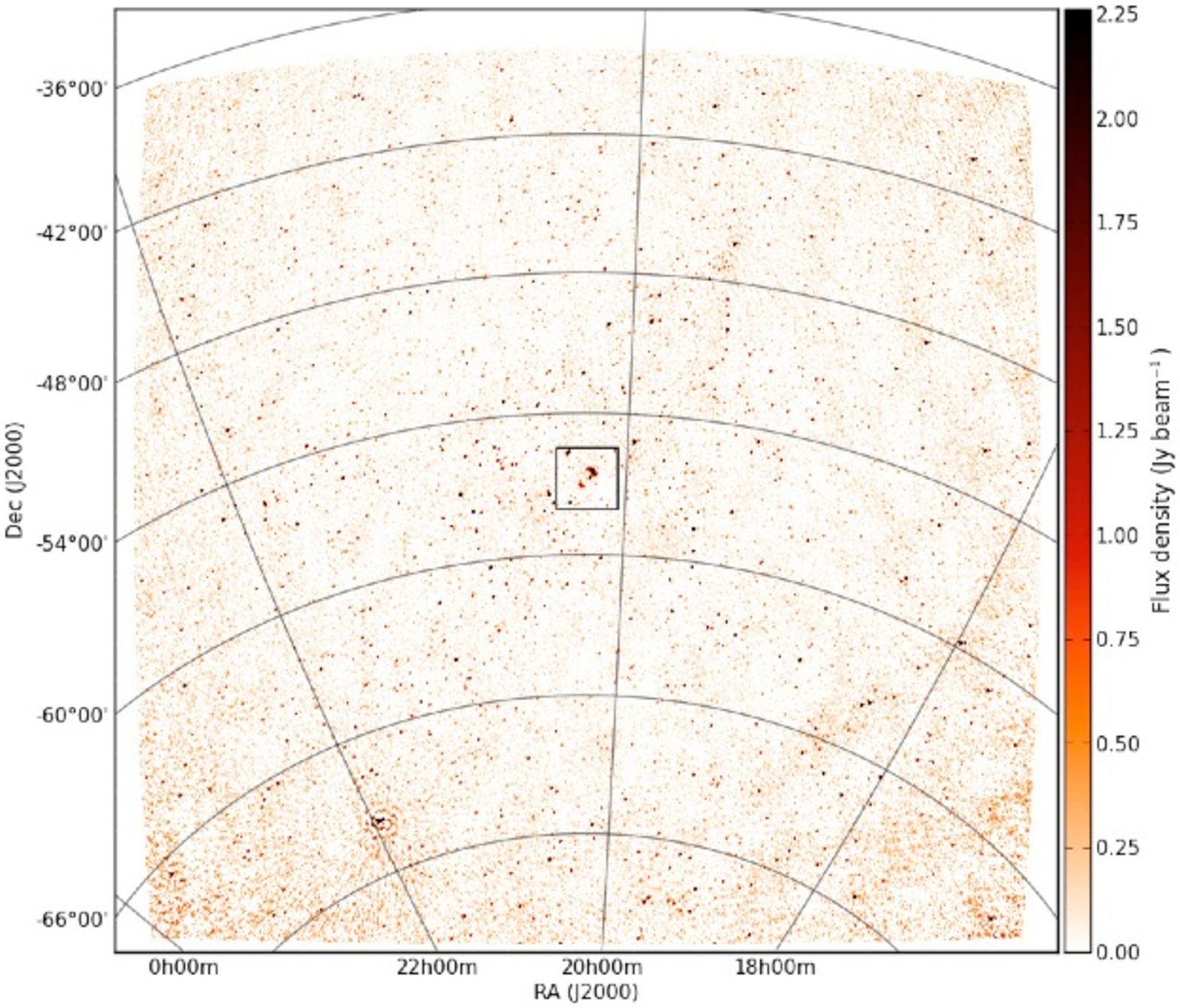} 
\includegraphics[width=0.49\textwidth]{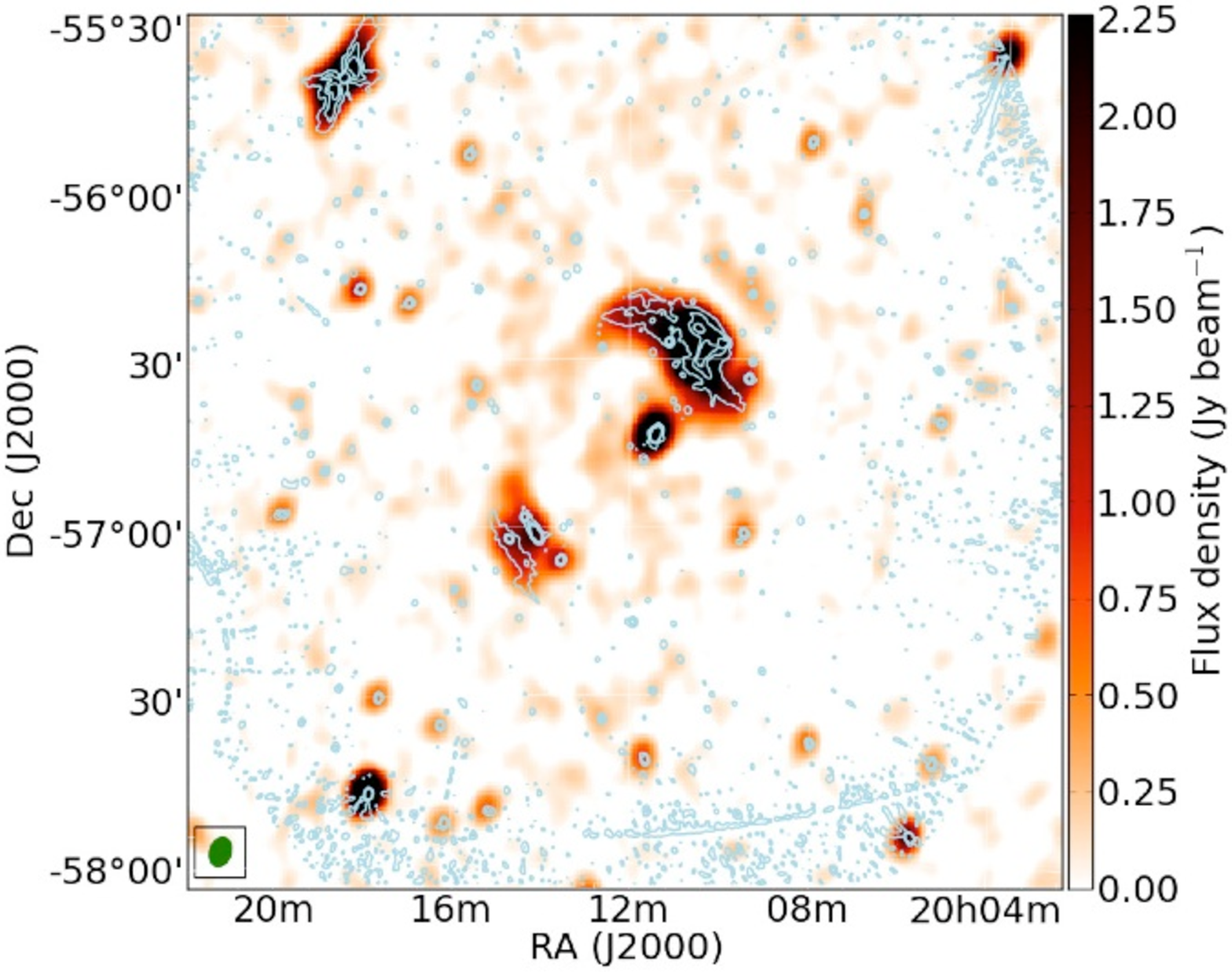} 
\includegraphics[width=0.49\textwidth]{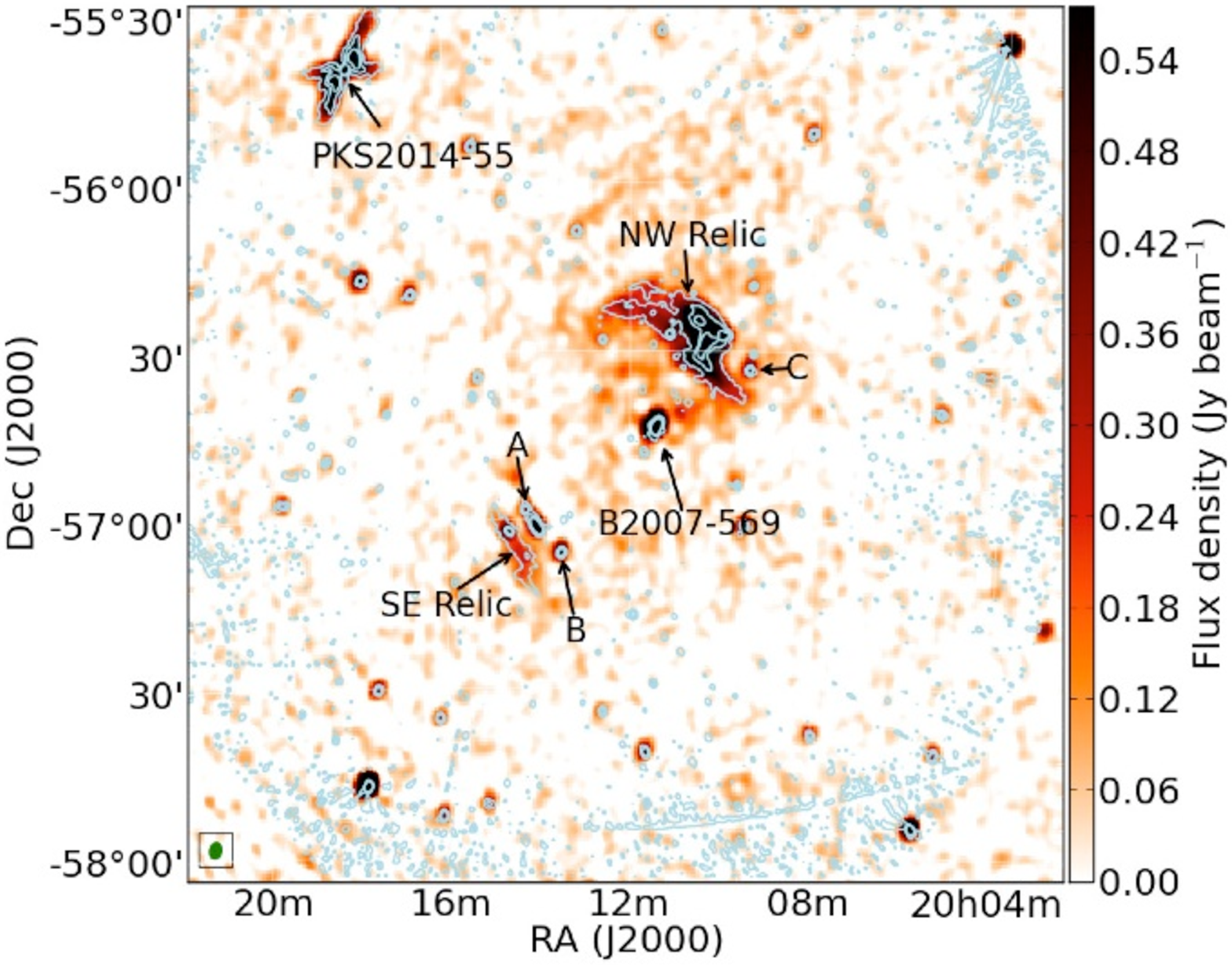} 
\caption{Top panel: full field of view of the primary beam and flux density corrected 120\,MHz image. The black box highlights A3667, which is shown in the bottom panels. Bottom panel: zoomed in images of A3667 at 120 and 226\,MHz (left and right, respectively). The synthesised beam is shown in the bottom left of each image and contours show 843\,MHz data taken with the MOST at 3\,mJy (5$\sigma$), 18 and 36\,mJy\,beam$^{-1}$. In the 226\,MHz image we label several features described in the text. The colour bar associated with each image shows the colour-scale in units of Jy.}
\label{im:radio}
\end{figure*}

In the top panel of Fig.~\ref{im:radio} we present the entire FoV of our 120\,MHz MWA image centred on A3667. This image highlights the incredibly wide FoV and high sensitivity obtained with just 12 minutes of integration. We present 120 and 226\,MHz images centered on A3667 with a smaller FoV in the bottom-left and bottom-right panels of Fig.~\ref{im:radio}. In these images we overlay contours taken from Molonglo Observatory Synthesis Telescope (MOST) observations as part of the SUMSS survey at 843\,MHz which have been reprocessed to remove artefacts and improve the sensitivity (see, \citealt{Hunstead1999,JohnstonHollitt2003} for processing details). This MOST image has a resolution of $0\farcm90\times0\farcm75$ and RMS sensitivity of $0.6$\,mJy\,beam$^{-1}$. Our MWA observations clearly detect the NW and SE relics on the periphery of the cluster. The peak surface brightnesses are $4.6\pm0.4$ and $1.1\pm0.1$\,Jy\,beam$^{-1}$ at 149\,MHz for the NW and SE relic, respectively. At a frequency of 120\,MHz we find that the SE relic becomes strongly blended with source\,A and to a lesser extent with source\,B. Source\,A (PMN\,J2014-5701) is a head-tail radio source with a bright core and narrow low surface brightness tail, with peak surface brightness $1.6\pm0.2$\,Jy\,beam$^{-1}$ at 149\,MHz. Source\,B (SUMSS\,J201330-570552) is a suspected FRII double-lobed radio galaxy \citep{JohnstonHollitt2003} and has a peak surface brightness of $0.9\pm0.1$\,Jy\,beam$^{-1}$ at 149\,MHz. Source\,C is an unresolved radio galaxy (SUMSS\,J200926-563328) with a peak surface brightness of $1.0\pm0.1$\,Jy\,beam$^{-1}$ which is blended with the NW relic at 120\,MHz. Towards the centre of the cluster, between the two relics, we identify the bright head-tail radio galaxy MRC\,B2007-568 \citep{Goss1982}, which has a peak surface brightness of $5.4\pm0.5$\,Jy\,beam$^{-1}$ at 149\,MHz. We also highlight the radio galaxy PKS\,B2014-55 that exhibits an interesting cross shaped morphology and find a peak brightness for this source of $4.1\pm0.4$\,Jy\,beam$^{-1}$.

\subsection{Integrated flux densities and spectral indices}\label{sect:int}

In order to characterise the extended emission in our MWA images we must first segment the emission into discrete sources. This is a non-trivial task when studying complex extended sources and requires the use of an automated process. We made use of the {\sc starlink}\footnote{http://starlink.jach.hawaii.edu/starlink/} tool {\sc fellwalker} \citep{Berry2007} to automatically assign the extended emission in each of the images into discrete regions. Algorithms such as {\sc fellwalker} are commonly used in the study of molecular gas in the Galaxy but the algorithms are equally valid for extended extragalactic radio emission. The {\sc fellwalker} algorithm works by finding the paths of steepest gradient from each pixel in an image above a user-defined threshold ($3\sigma$). Starting with the first pixel in the image, each of the surrounding pixels are inspected to locate the pixel with the highest ascending gradient; this process continues until a peak is located (i.e.~a pixel surrounded by flat or descending gradients). The pixels along this path are assigned an arbitrary integer to represent their connection along a path. All pixels in the image are inspected and the image is segmented into clumps by grouping together all paths that lead to the same peak value. The pixels belonging to paths that lead to a single peak are then defined as belonging to that particular source. For a complete description of this process see, \citet{Berry2007}.

We determine the integrated flux density of the NW and SE relics, MRC\,B2007-568, PKS\,B2014-55 and sources\,A, B and C using the output of {\sc fellwalker} as a mask and extracting the integrated flux density for each source at each frequency band in our MWA and MOST images. There are a number of sources in the field that contaminate the flux density estimates for the NW and SE relics. Some of these sources are obvious in our MWA images, such as source\,A, which is blended with the SE relic and Source\,C, which is blended with the NW relic in the 120\,MHz image (Fig.~\ref{im:radio}; bottom-left). The mask generated by {\sc fellwalker} is unable to separate source\,A from the SE relic or source\,C from the NW relic. We therefore estimate the integrated flux density for source\,A and C at 120\,MHz by extrapolating the spectral profile using the spectral index derived from flux density measurements above 120\,MHz. We then subtract the estimated integrated flux density for sources\,A and C from our integrated flux density estimates for the NW and SE relics at 120\,MHz.

There are also a number of point sources that are coincident with the NW and SE relics, which are only visible in the 843\,MHz MOST image (Fig.~\ref{im:radio}; bottom panels). These point sources are blended with the NW and SE relics at all frequencies in our MWA images and so will contaminate the integrated flux density estimates. We find 16 point sources in the 843\,MHz MOST image that are associated with the NW relic and two associated with the SE relic. These point sources have a combined flux density of 529.6 and 63.6\,mJy at 843\,MHz for the NW and SE relics, respectively. In order to account for the flux density contributed by these point sources we extrapolate the combined flux density from the 843\,MHz image to the appropriate MWA central frequencies assuming a representative spectral index for radio galaxies of $-0.7\pm0.2$. In this way we find the total flux density of the contaminating point sources to be $\sim 1.8$ and $\sim 0.2$\,Jy at 149\,MHz for the NW and SE relic, respectively. Ideally we would like to model and peel these sources from the dataset. Unfortunately, there are no reliable data at the same or similar frequency with high enough resolution to accurately model these sources. To remove these contaminating point sources we simply subtract their combined total flux density from our flux density estimates for the NW and SE relics.

\begin{figure*}
\centering
\includegraphics[width=0.4\textwidth]{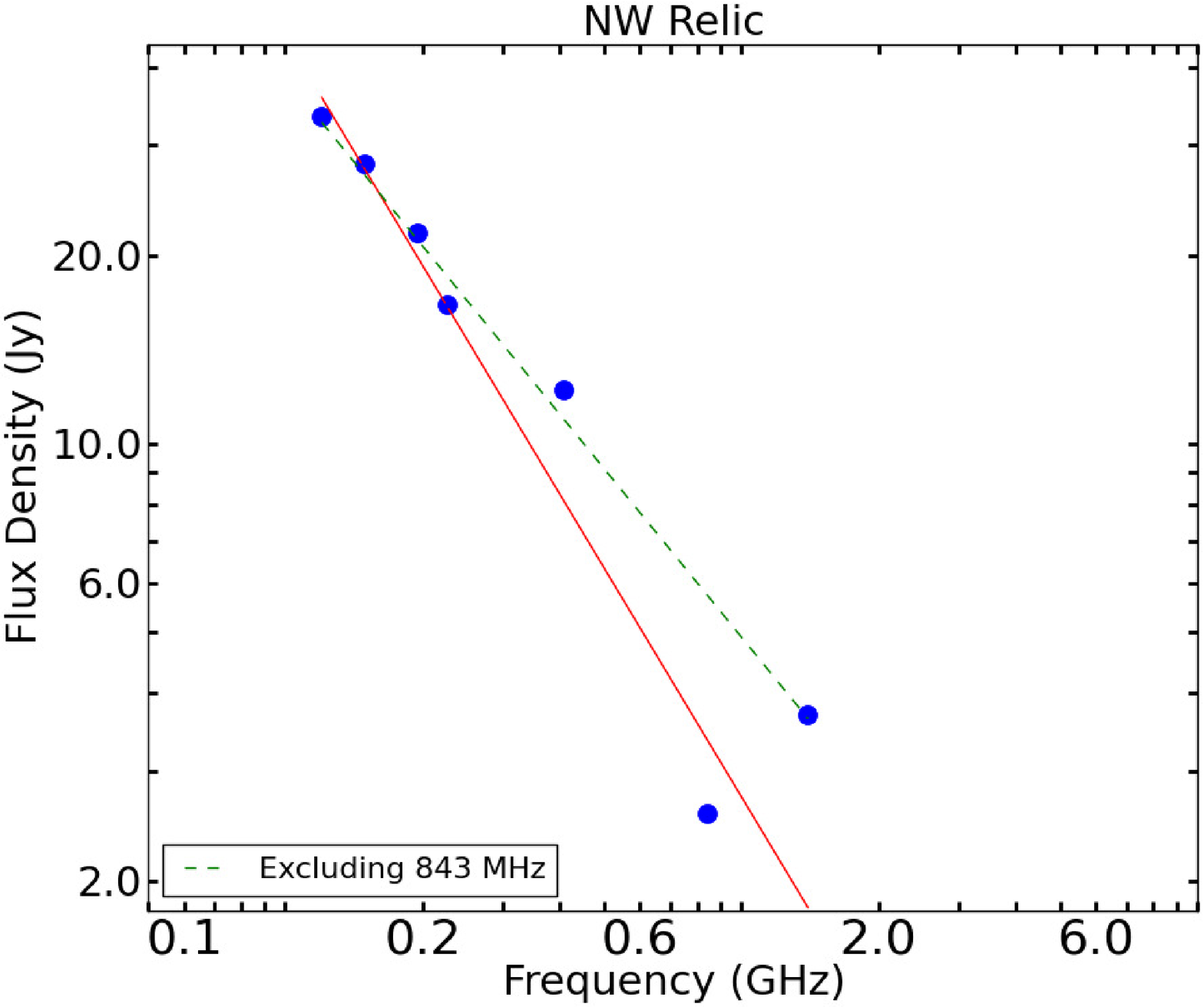} 
\includegraphics[width=0.4\textwidth]{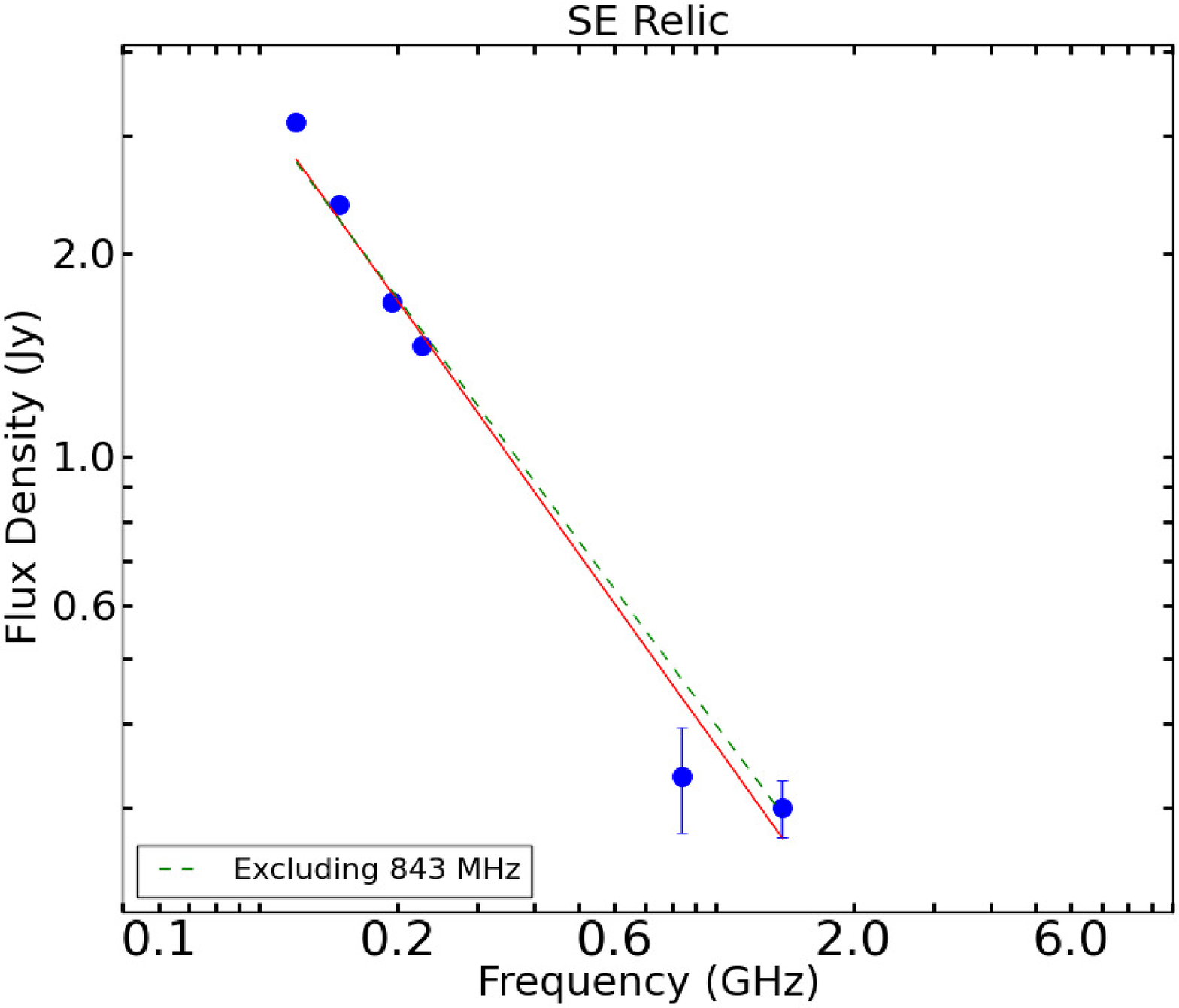} 
\includegraphics[width=0.4\textwidth]{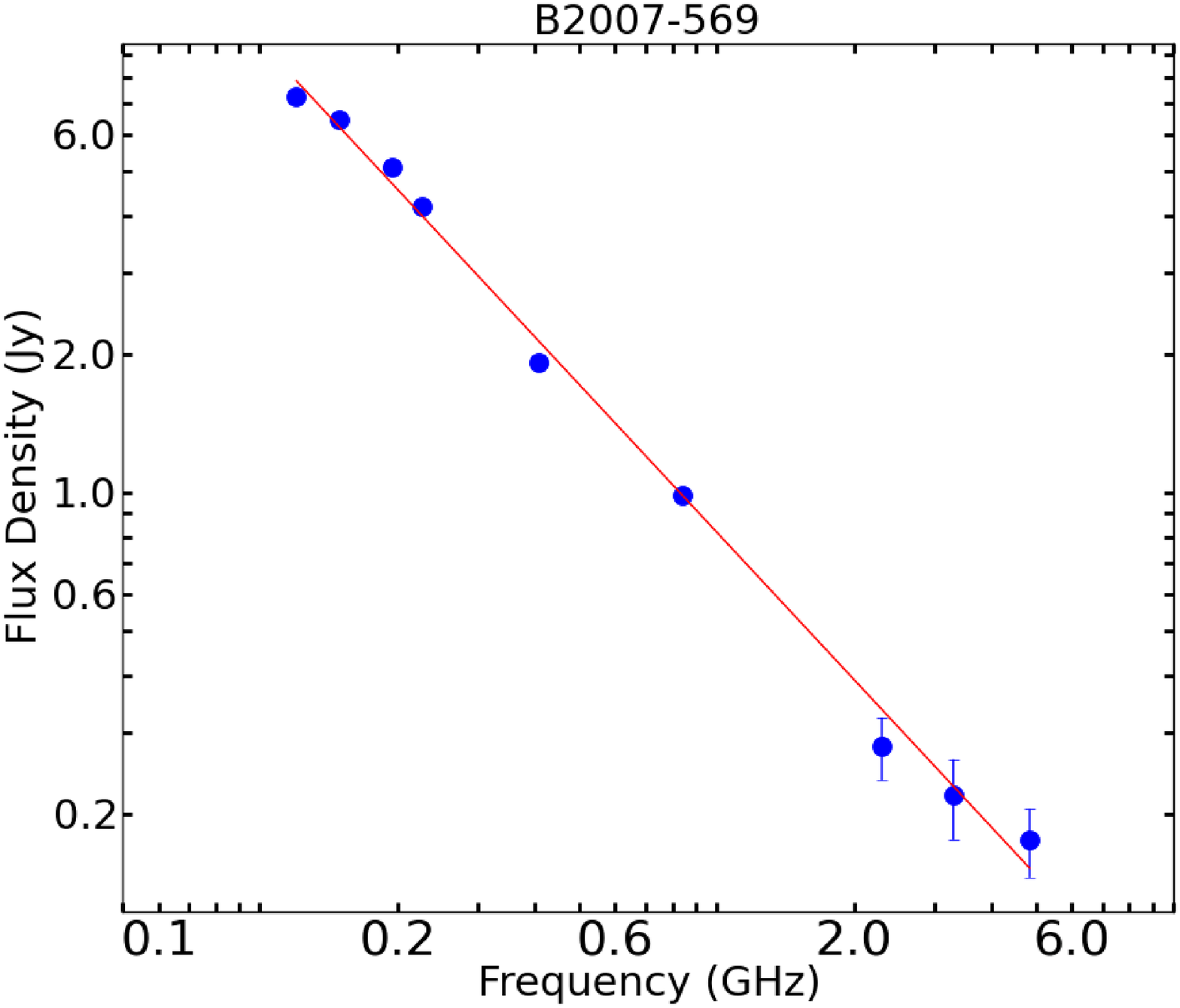} 
\includegraphics[width=0.4\textwidth]{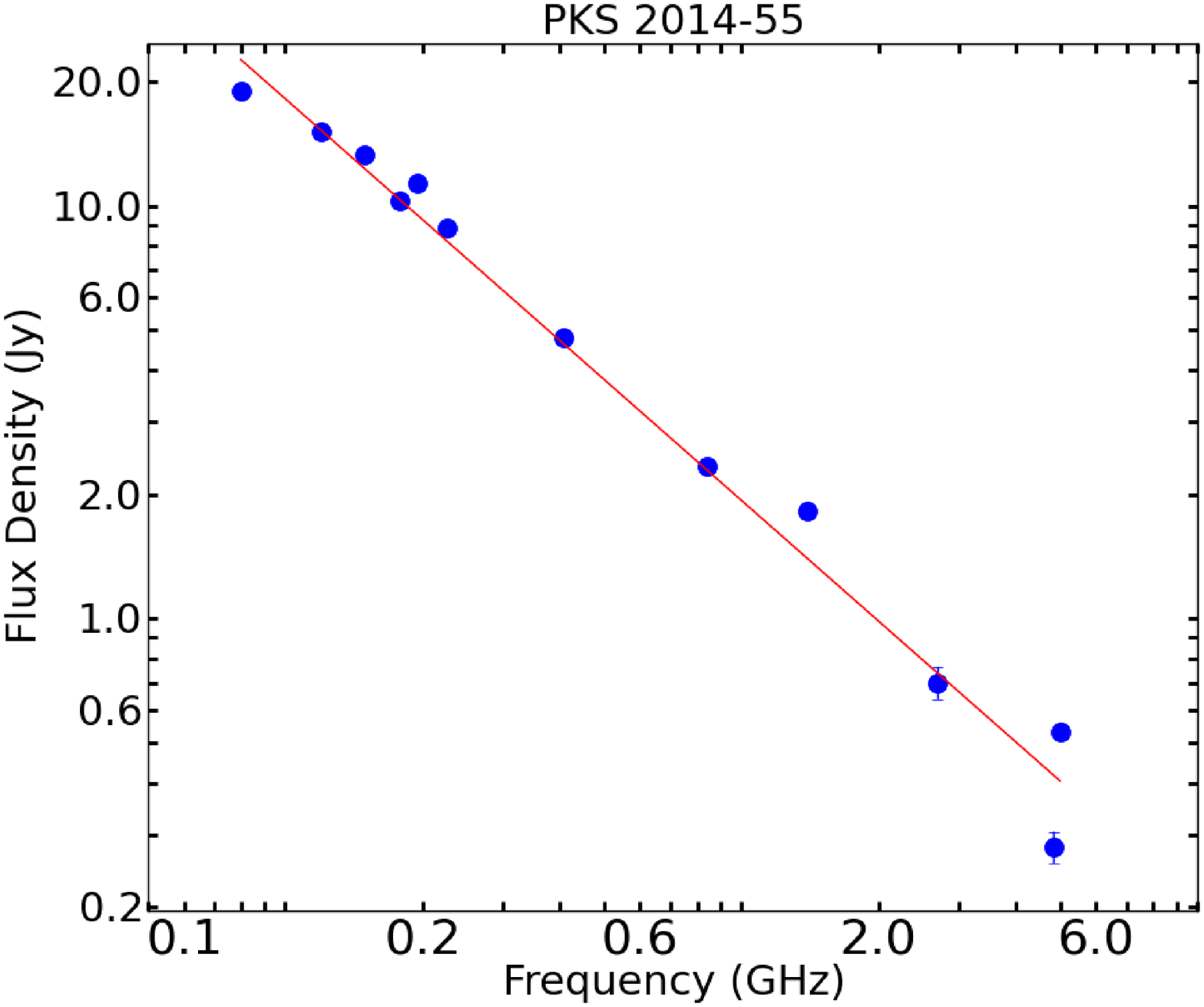} 
\includegraphics[width=0.4\textwidth]{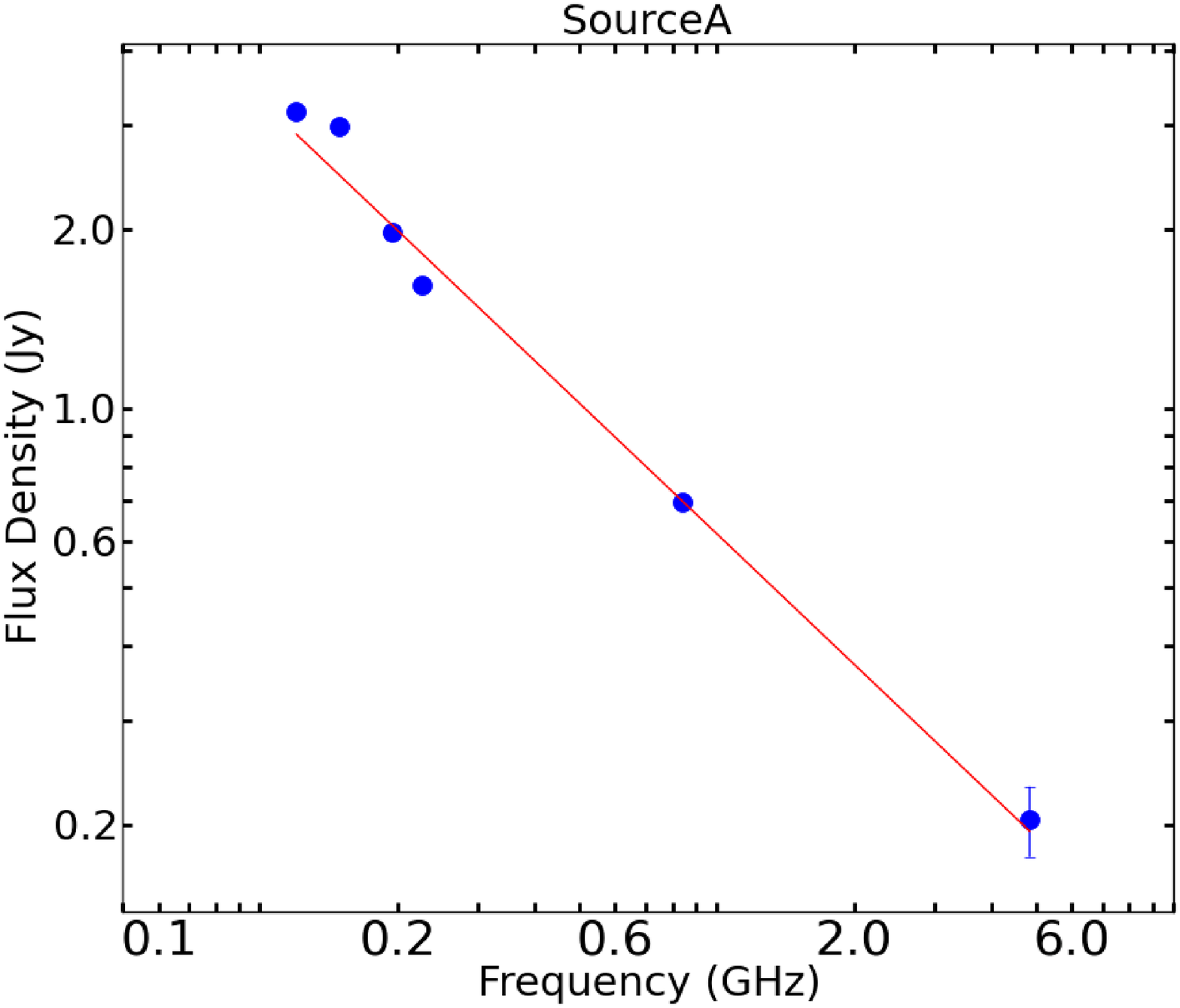} 
\includegraphics[width=0.4\textwidth]{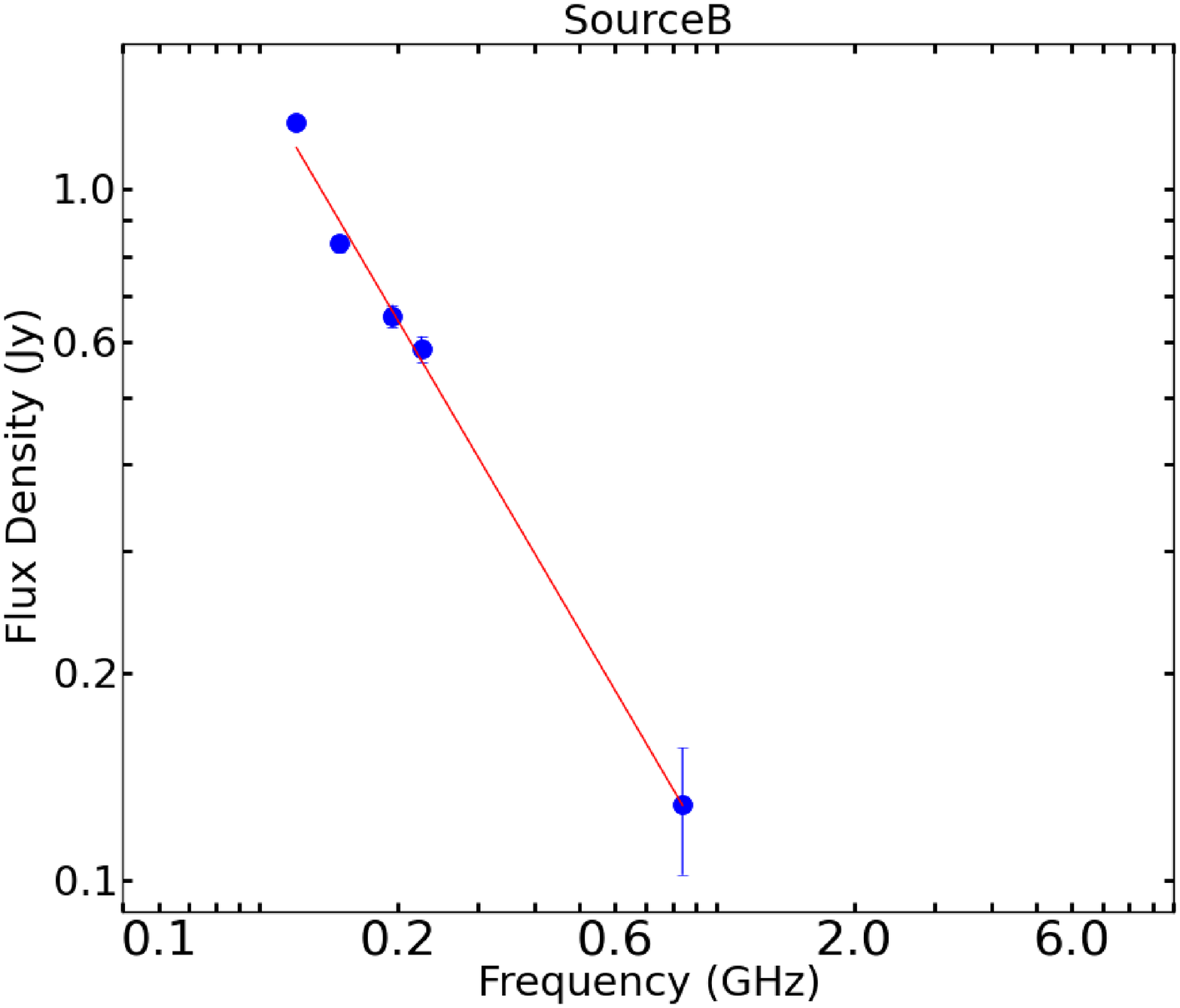} 
\includegraphics[width=0.4\textwidth]{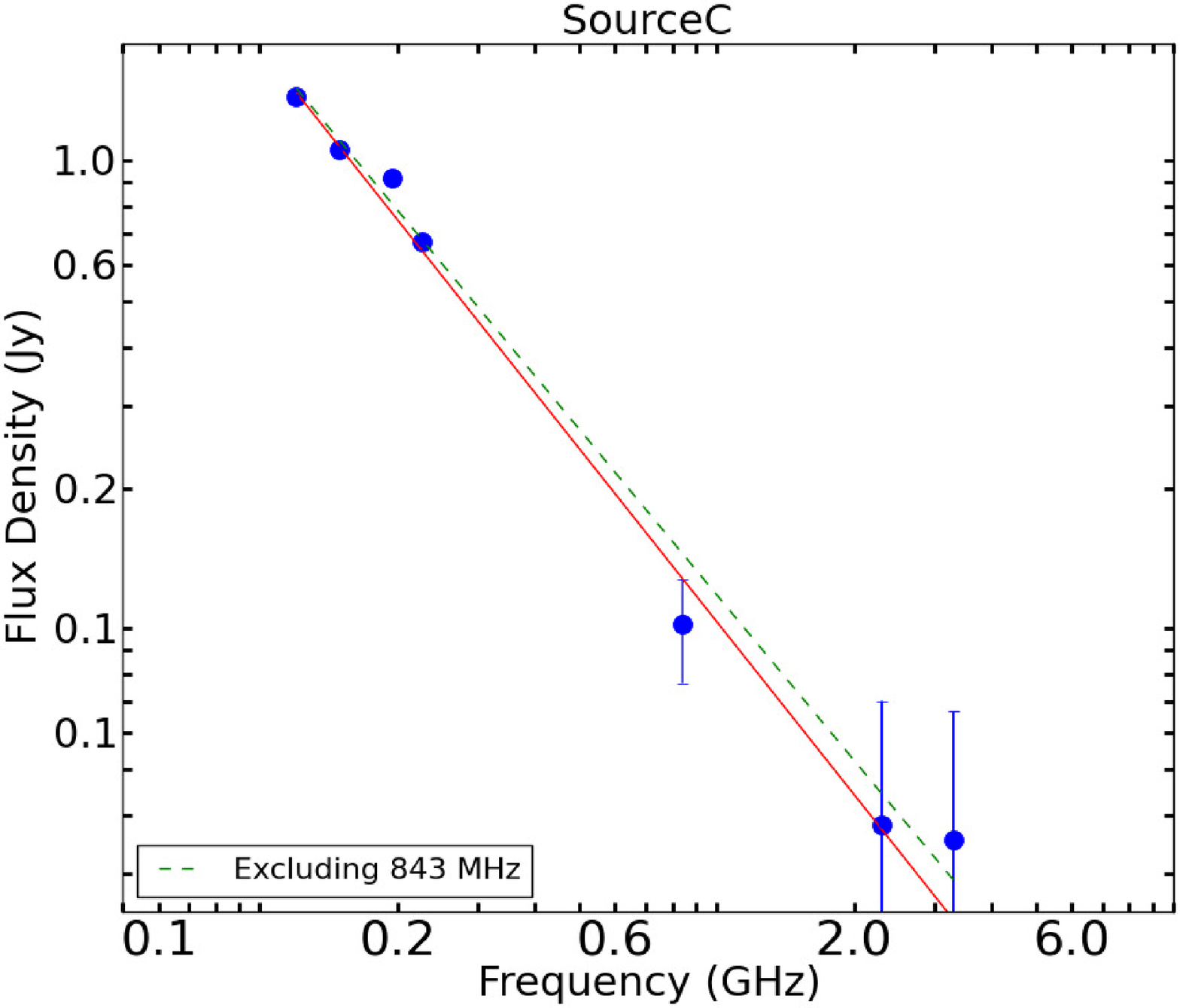} 
\caption{Spectral energy distributions of the sources of interest labelled in Fig.~\ref{im:radio}. We fit a power-law (red line) to the spectral profile using flux density measurements from our MWA images (Table~\ref{tab:props}), MOST data and the literature (Table~\ref{tab:literature}). We include a fit to the spectral profile excluding the 843\,MHz point for the NW and SE relics and Source\,C (green dashed line).}
\label{im:intfit}
\end{figure*}

We derive the spectral index from the integrated flux densities by fitting a power-law between the MWA bands, MOST data and flux density measurements taken from the literature between 86 and 4850\,MHz (Table~\ref{tab:literature}). In this way we find an average spectral index for the NW relic, between 86 and 1400\,MHz, of $-1.2\pm 0.1$. For the SE relic we find a spectral index, between 120 and 1400\,MHz, of $-0.9\pm0.1$ (Fig.~\ref{im:intfit}; red line). We find the spectral index derived for the NW and SE relics are consistent, within the errors, with higher frequency and resolution data between 0.8 and 1.4\,GHz, where the spectral index for the NW and SE relics are $-0.9\pm0.2$ and $-1.2\pm0.2$ \citep{JohnstonHollitt2003}, respectively. However, we note that the reduced $\chi^2$ value for our fits to the spectral profile of the NW and SE relic is high (143.6 and 45.8, respectively) indicating a poor fit.

Inspection of the spectral profiles in Fig.~\ref{im:intfit} reveals that this poor fit is caused by the integrated flux density measurement at 843\,MHz, which appears to be underestimated for both of the relics in A3667 and Source\,C (Fig.~\ref{im:intfit}). If we remove the flux density at 843\,MHz from our fitting procedure we find that the NW relic has a flatter spectral index of $-0.9\pm0.1$ (Fig.~\ref{im:intfit}; dashed-green line) with reduced $\chi^2$ value of 5.2. The spectral index for the SE relic and Source\,C remains the same within the quoted errors. Sources A, B, MRC\,B2007-568 and PKS\,B2014-55 are all isolated and fairly compact regions of emission and have the same spectral index, within the errors, with and without the 843\,MHz flux density. The NW and SE relics and Source\,C are associated with extended low-surface brightness emission. We suspect that spatial filtering caused by missing short-spacings in the MOST observations is resulting in missing flux associated with emission on the scale of the NW relic. There is a 15-m gap at zero-spacing between the two arms of the MOST \citep{Mills1981}, which results in emission on angular scales of $\sim 30\arcmin$ and above being poorly sampled. Our results agree with this: the 843\,MHz data is able to recover the total flux density for compact sources but for emission that is on the scale of the NW relic ($\sim30\arcmin$) there is a significant amount of missing flux. Given the shortest baseline of 7.7\,m we do not expect our MWA images to be missing a significant amount of flux below scales of $600\arcmin$. We present the integrated flux density and spectral indices for the NW and SE relics, MRC\,B2007-568, PKS\,B2014-55 and sources A, B and C in Table~\ref{tab:props}.

\begin{table}
 \caption{Flux density measurements of sources labelled in Fig.~\ref{im:radio}.}
\begin{tabular}{lccr}
 \hline 
 Source & Frequency & Flux Density & Reference \\
 		Name & (MHz) & (Jy) &   \\		
 	\hline
 	
NW Relic & 86 & $81.0\pm8.0$ & \cite{Mills1961}   \\
 & 408 & $12.1\pm1.2$ & \cite{Bolton1964}  \\
 & 843 & $2.6\pm0.2$ & This work  \\
 & 1400 & $3.70\pm0.30$ &  \cite{JohnstonHollitt2003} \\
SE Relic & 843 & $0.33\pm0.05$ & This work  \\
 & 1400 & $0.30\pm 0.02$ &  \cite{JohnstonHollitt2003}  \\
MRC\,B2007-568 & 408 & $1.92\pm0.19$ & \cite{Large1991} \\
 & 843 & $1.0\pm0.1$ & This work  \\
 & 2300 & $0.28\pm0.03$ & \cite{Carretti2013} \\
 & 3300 & $0.22\pm0.02$ & \cite{Carretti2013} \\
 & 4850 & $0.18\pm0.01$ & \cite{Wright1994} \\ 
Source\,A & 843 & $0.7\pm0.1$ & This work  \\
 & 4850 & $0.20\pm0.01$ &\cite{Wright1994} \\
Source\,B & 843 & $0.13\pm0.01$ & This work  \\
Source\,C  & 843 & $0.10\pm0.01$ & This work  \\
 & 2300 & $0.038$ &\cite{Carretti2013} \\
 & 3300 & $0.036$ & \cite{Carretti2013} \\
PKS\,B2014-55 & 86 & $19.0\pm0.1$ & \cite{Mills1961} \\
 & 178 & $10.3\pm0.1$ & \cite{Mills1961} \\
 & 408 & $4.8\pm0.3$ & \cite{Schilizzi1975} \\
 & 843 & $2.3\pm0.1$ & This work  \\
 & 1400 & $1.82\pm0.20$ & \cite{Smith1983} \\
 & 2700 & $0.7\pm0.1$ & \cite{Bolton1964} \\
 & 4850 & $0.28\pm0.02$ & \cite{Wright1994} \\
 & 5000 & $0.28\pm0.02$ & \cite{Wall1979} \\
 \hline
 \end{tabular}
 \label{tab:literature}
 \end{table}

\subsection{T-T plot}

\begin{figure}
\includegraphics[width=0.5\textwidth]{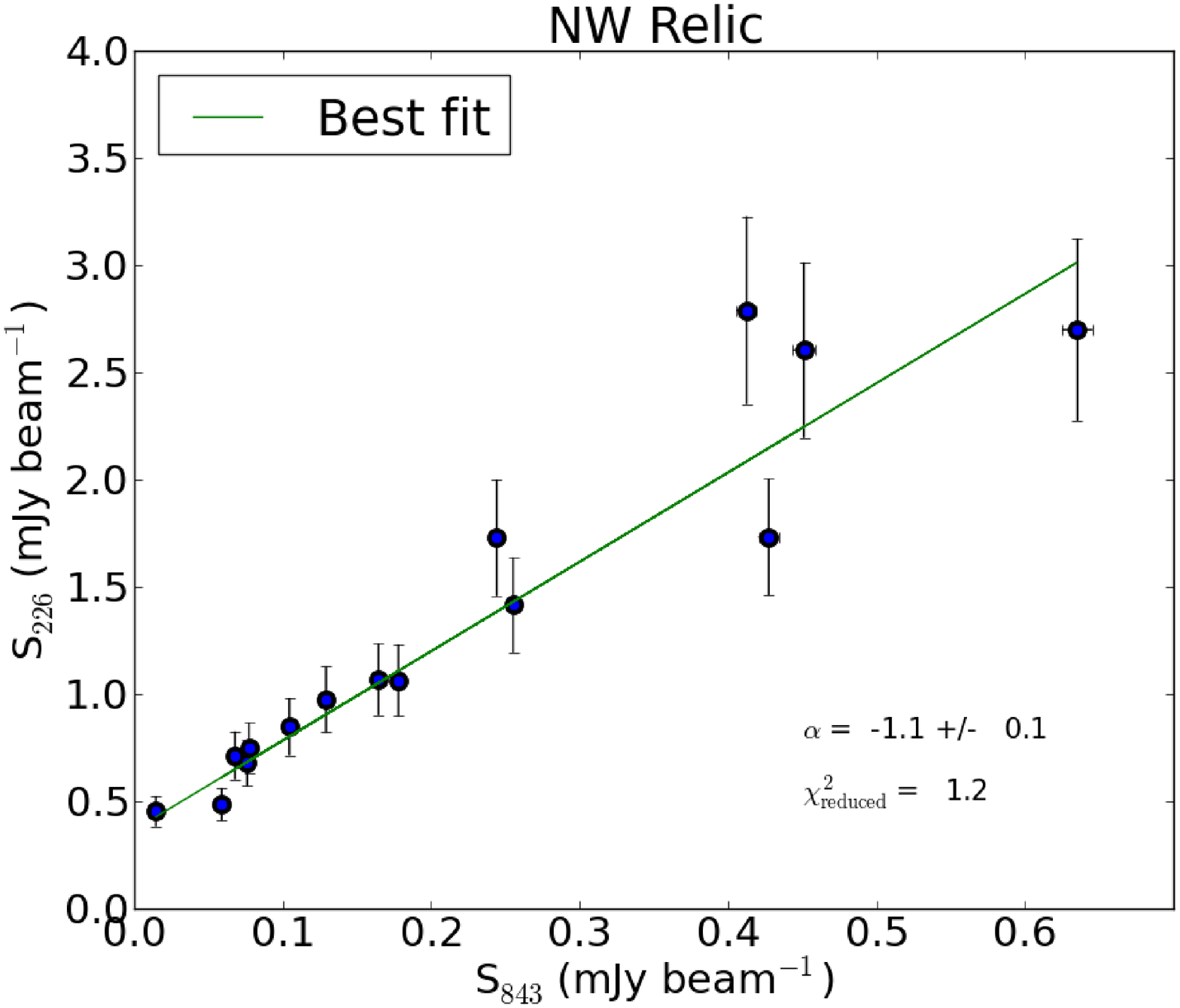} 
\caption{T-T plot for the NW relic generated from the 226 and 843\,MHz images. The images have been smoothed and regridded so a single pixel corresponds to the size of the beam. The fitted line indicates the spectral index of the region.}
\label{im:tt}
\end{figure}

To investigate whether there is any spatial variation of the spectral index across the NW relic and search for further evidence of missing flux in our 843\,MHz image we construct a T-T plot \citep{Turtle1962} using the 226 and 843\,MHz images in a similar manner to \cite{mckinley2013}. We are unable to generate a T-T plot for the SE relic due to the low resolution and small angular size of the relic. T-T plots are a useful tool for comparing the flux density at two frequencies from different instruments because they are able to account for the possibility of variations in the short spacings and possible foreground contamination. Any difference in the short spacings will manifest as a non-zero y-axis intercept in T-T plot. To generate a T-T plot between the 843\,MHz and 226\,MHz images we first smooth the 843\,MHz image to match the 226\,MHz resolution. We then regrid both images so that a single pixel corresponds to the size of the synthesised beam. In this way we generate two images with independent flux density measures, allowing us to perform $\chi^2$ analysis. The resulting plot of 843 vs. 226\,MHz flux density for each pixel in the NW relic can be seen in Fig.~\ref{im:tt}. By fitting a straight line to the points on this plot, using orthogonal-distance regression, we extract a spectral index for the region assuming a constant spectral index over the source. Using this method we find a spectral index of $-1.1\pm0.1$ which is consistent with our integrated flux estimates in the previous section. Computing the reduced $\chi^{2}$ value of the fitted line reveals a value of 1.2, which suggests that there is some minor spatial variation in the spectral index across the NW relic at the resolution of these observations. This is consistent with higher resolution and frequency observations which show a variation in the spectral index which indicates that more than a single electron population is responsible for the observed emission \citep{JohnstonHollitt2003}. The y-axis intercept of the line fitted to the T-T plot is 0.3 which indicates that there is indeed an offset between the flux density detected in the 226\,MHz image and the 843\,MHz MOST image.

\begin{table*}
\setlength{\tabcolsep}{5pt}
 \caption{Observed properties of the features in the MWA images. The $^*$ above the integrated flux density at 120\,MHz for sources A and C indicates that this integrated flux density could not be extracted directly but has been extrapolated by fitting a power-law to the higher frequency data. The integrated flux densities for the NW and SE relics have had the flux density of contaminating sources subtracted (see text for details). The spectral indices are determined by fitting a power-law to the flux density in each of the MWA bands, MOST data and data taken form the literature presented in Table~\ref{tab:literature}. The spectral index fitted for the NW relic does not include the 843\,MHz measurement.}
\begin{tabular}{l|cc|cccc|cccc|c}
 \hline 
 Source & \multicolumn{2}{c|}{Peak Position} & \multicolumn{4}{c|}{Peak Surface Brightness (Jy\,beam$^{-1}$)} & \multicolumn{4}{c|}{Integrated Flux Density (Jy)} & Spectral \\
 Name & RA & DEC & 120\,MHz & 149\,MHz & 180\,MHz & 226\,MHz & 120\,MHz & 149\,MHz & 180\,MHz& 226\,MHz & Index  \\ 
 \hline
NW Relic   & $  302.52  $ & $  -56.53  $ & $  7.5  \pm  0.7  $ & $  4.6  \pm  0.4  $ & $  2.7  \pm  0.2  $ & $  1.8  \pm  0.2  $ & $  33.4  \pm  2.0  $ & $  28.1  \pm  1.7  $ & $  21.8  \pm  1.3  $ & $  16.7  \pm  1.0  $ & $  -0.9  \pm  0.1 $\\
SE Relic   & $  303.54  $ & $  -57.12  $ & $  2.1  \pm  0.2  $ & $  1.1  \pm  0.1  $ & $  0.7  \pm  0.1  $ & $  0.6  \pm  0.1  $ & $  3.2  \pm  0.4  $ & $  2.4  \pm  0.1  $ & $  1.7  \pm  0.1  $ & $  1.5  \pm  0.1  $ & $  -0.9  \pm  0.1  $\\
MRC\,B2007-568   & $  302.97  $ & $  -56.68  $ & $  6.7  \pm  0.6  $ & $  5.4  \pm  0.5  $ & $  4.1  \pm  0.3  $ & $  3.6  \pm  0.4  $ & $  7.3  \pm  0.4  $ & $  6.5  \pm  0.4  $ & $  5.1  \pm  0.3  $ & $  4.2  \pm  0.2  $ & $  -1.1  \pm  0.1  $ \\
Source\,A   & $  303.51  $ & $  -57.03  $ & $  2.1  \pm  0.2  $ & $  1.6  \pm  0.2  $ & $  1.2  \pm  0.1  $ & $  1.0  \pm  0.1  $ & $  3.2^{*}  \pm  0.3  $ & $  3.0  \pm  0.2  $ & $  2.0  \pm  0.1  $ & $  1.6  \pm  0.1  $ & $  -0.7  \pm  0.1  $ \\
Source\,B   & $  303.65  $ & $  -56.97  $ & $  1.3  \pm  0.1  $ & $  0.9  \pm  0.1  $ & $  0.8  \pm  0.1  $ & $  0.7  \pm  0.1  $ & $  1.2  \pm  0.1  $ & $  0.8  \pm  0.1  $ & $  0.7  \pm  0.1  $ & $  0.6  \pm  0.1  $ & $  -1.1  \pm  0.1  $ \\
Source\,C   & $  302.67  $ & $  -56.40  $ & $  1.6^{*}  \pm  0.2  $ & $  1.0  \pm  0.1  $ & $  0.7  \pm  0.1  $ & $  0.5  \pm  0.1  $ & $  1.4  \pm  0.1  $ & $  1.1  \pm  0.1  $ & $  0.9  \pm  0.1  $ & $  0.7  \pm  0.1  $ & $  -1.2  \pm  0.1  $\\
PKS\,2014-55   & $  300.98  $ & $  -57.58  $ & $  5.8  \pm  0.5  $ & $  4.1  \pm  0.4  $ & $  2.7  \pm  0.2  $ & $  1.9  \pm  0.2  $ & $  15.2  \pm  0.8  $ & $  13.3  \pm  0.7  $ & $  11.3  \pm  0.6  $ & $  8.8  \pm  0.5  $ & $  -1.0  \pm  0.1 $ \\
 \hline
 \end{tabular}
 \label{tab:props}
 \end{table*}

\subsection{Spectral index maps}

\begin{figure}
\includegraphics[width=0.5\textwidth]{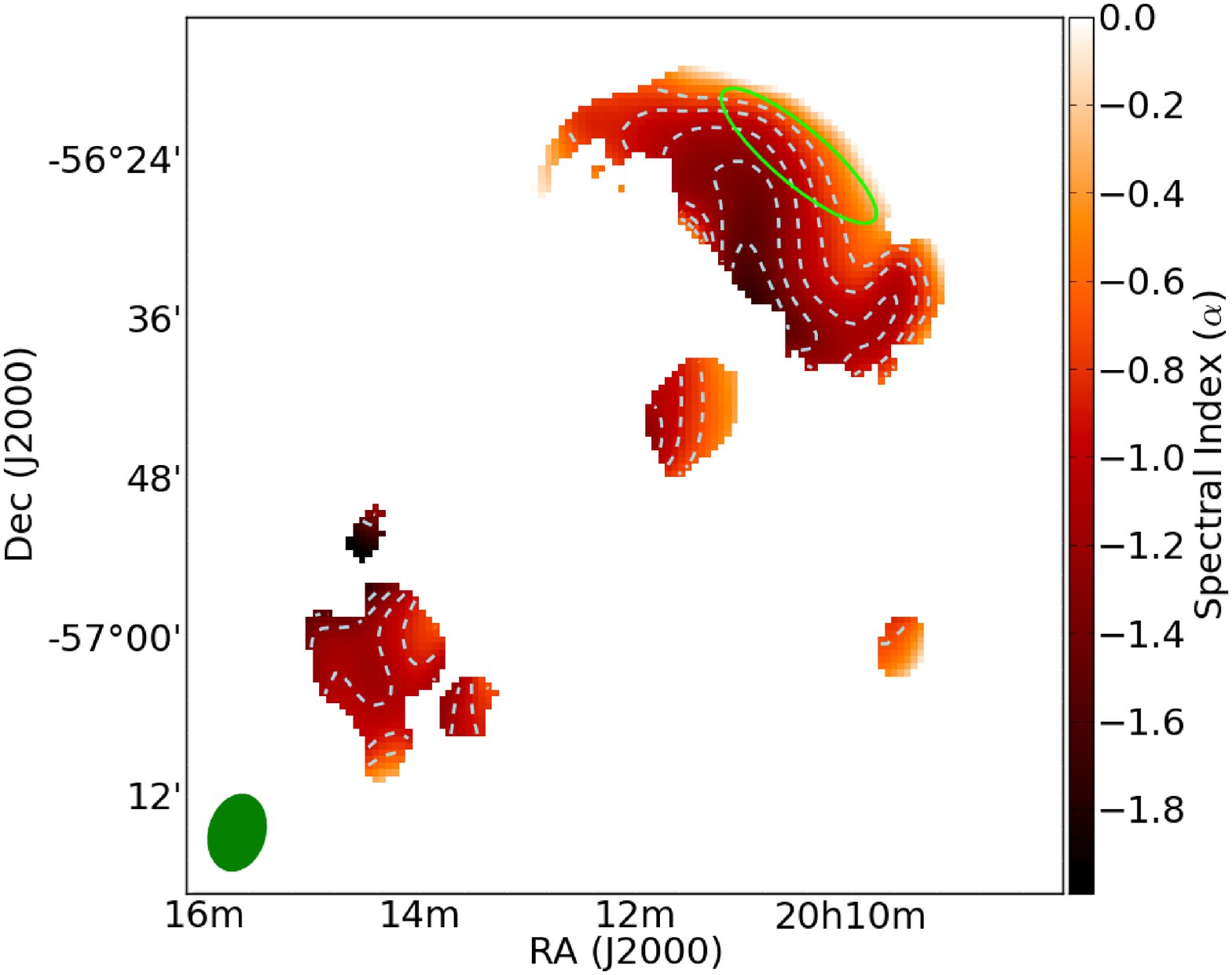} 
\includegraphics[width=0.5\textwidth]{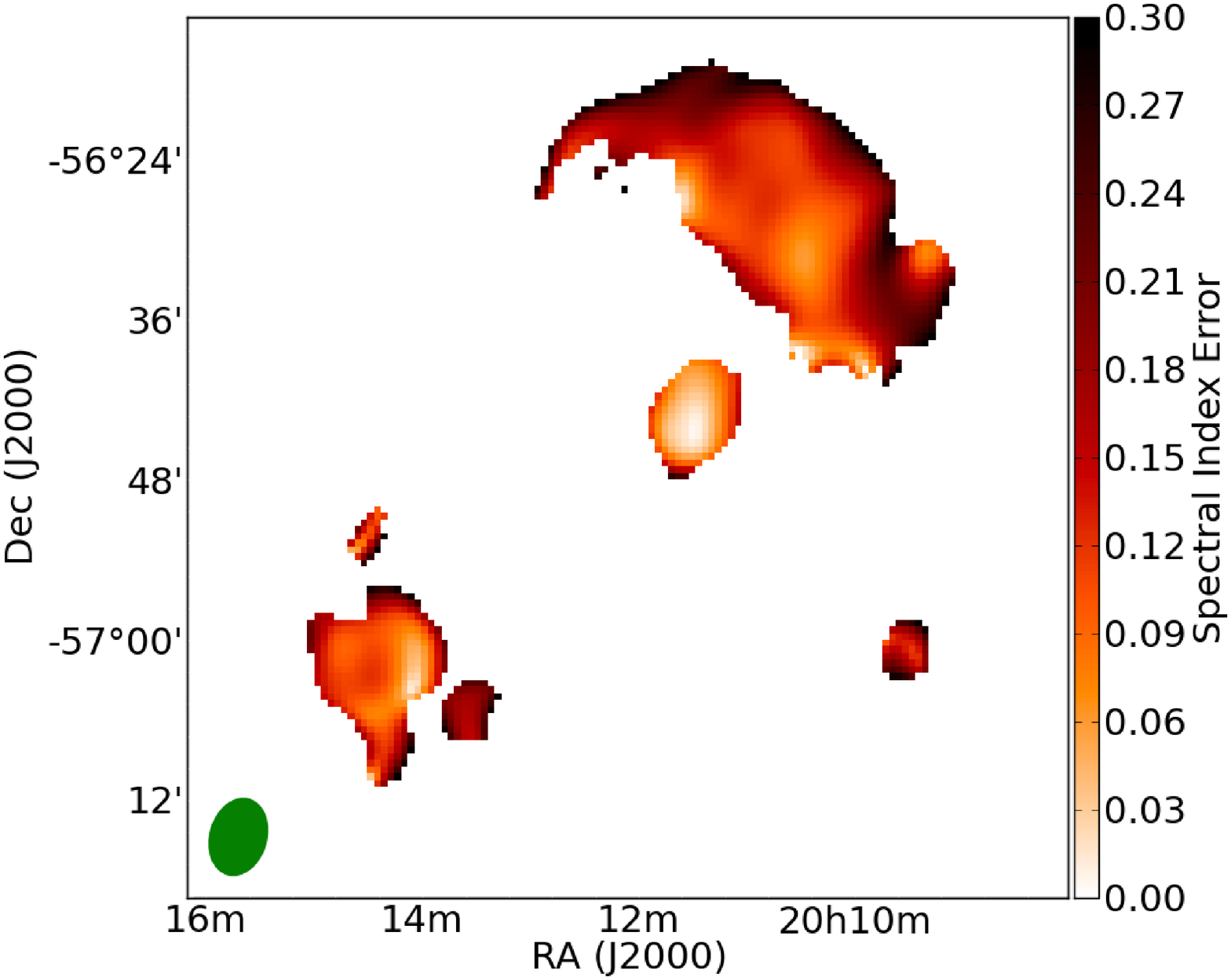} 
\caption{Top panel: spectral index map generated from a four point power-law fit to the flux density for each pixel in the four MWA band images. Light blue contours show the spectral index from $-1.5$ to $-0.7$ incrementing by $0.2$. A green ellipse indicates the area over which we determine the spectral index of the leading edge of the NW relic. Bottom panel: the associated error in the fitted spectral index found by a linear least-squares regression. The smoothed beam size is shown in the bottom-left.}
\label{im:specindex}
\end{figure}

Given the missing flux in our 843\,MHz MOST image, we generate a spectral index map of A3667 between 120 and 226\,MHz. We first convolve each of the MWA images to a matching resolution that is 10\% larger than the beam at 120\,MHz. This ensures that we smooth each image to the same resolution and so sample each pixel equally. We then fit a four-point power-law to each pixel in the image using a least-squares routine. We do not apply any flux density cut-off when generating the initial spectral index map to avoid any bias associated with flat spectrum emission that would be present in the high frequency data but below $3\sigma$ in the lower frequency images. We apply our {\sc fellwalker} mask to the resultant spectral index map which can be seen in the top panel of Fig.~\ref{im:specindex} along with the associated error of the fitted spectral index in the bottom panel of Fig.~\ref{im:specindex}. This error is the standard deviation of the fitted spectral index and includes both the RMS uncertainty in each image as well as the uncertainty derived from our flux calibration step.  We generate a map of believable spectral indices by masking spectral index measurements that have an associated error that is greater than $0.3$.

The spectral index between 120 and 226\,MHz in these maps ranges from $-1.7\pm0.2$ to $-0.4\pm0.2$ with a pixel average of $-1.0\pm0.2$ for the NW relic. For the SE relic we find an average spectral index between 120 and 226\,MHz of $-1.1\pm0.2$ with a range of $-1.5\pm0.2$ to $-0.4\pm0.2$. These values for the average spectral index are consistent with the spectral indices presented in Table~\ref{tab:props}. We are able to clearly detect some variation in the spectral index across the NW relic. In particular the central region of the NW relic shows variation on large scales that is consistent with the spectral index between 0.8 and 1.4\,GHz \citep{JohnstonHollitt2003}. We find evidence of a spatial gradient in the spectral index from the leading edge of the NW relic towards the centre of the cluster. The spectral index at the leading edge of the relic is approximately $-0.4\pm0.2$ and steepens to $-1.7\pm0.2$ towards the centre of the cluster. This indicates that there may be a change in the age of the electron population responsible for the observed radio emission. For the SE relic we are unable to detect a gradient in the spectral index transverse to the shock front seen at higher frequencies \citep{JohnstonHollitt2003}. This is due to the low resolution of these MWA observations. 

\section{Discussion}\label{Sect:discussion}

\subsection{Radio relics in A3667}\label{SubSect:relics}

The spatial distribution of the spectral index across the NW relic is non-uniform. This has been reported in high resolution observations at 1.4\,GHz where the relic is observed to have a highly complex filamentary structure \citep{JohnstonHollitt2003}. We suspect that the NW relic is not being viewed edge-on and/or has a complex geometry, which makes subsequent analysis difficult. The NW relic has a length and width of approximately 1.9 by 0.9\,Mpc, respectively. If we were viewing the NW relic edge-on we would expect a much larger ratio between the length and width such as is seen in the toothbrush relic and the relics in CIZA J2242.8+5301 \citep{VanWeeren2012,Stroe2013}. If we assume that the relic traces a planar shock located in the xy-plane (i.e the plane of the sky), has a similar extent in the x and y direction and a negligible extent in the downstream region we can estimate an upper limit for the viewing angle of $\sim 26\degr$ from edge-on. This geometry will have implications on any analysis of the spectral index but is difficult to quantify. We are unable to resolve substructure in the SE relic and so concentrate our discussion on the NW relic.

The spectral index of the NW relic has a clear gradient; the spectral index is flattest at the leading edge, with a value of $-0.4\pm0.2$, and steepens downstream towards the cluster centre peaking with a value of $-1.7\pm0.2$. If we ignore projection effects then such a gradient is predicted by the diffusive shock acceleration (DSA) mechanism whereby merging sub-clusters generate shocks in the ICM, which propagate outwards from the cluster centre \citep{Blandford1987,Bonafede2014,Bonafede2012,Bonafede2009,VanWeeren2012A,VanWeeren2010}. The acceleration of electrons due to DSA results in a power-law spectrum of the accelerated electrons. Assuming a fully ionised plasma, the slope of the electron power-law spectrum, $\delta$, is related to the Mach number, $M$, of the shock by \citep{Blandford1987}:

\begin{equation}
\delta=-2\frac{M^2+1}{M^2-1},
\end{equation}

\noindent where the spectral index of the synchrotron emission associated with the accelerated electrons is given by $\alpha = (\delta +1)/2$. If we assume that the shock is located at the leading edge of the NW radio relic, which accelerates a single electron population and that the synchrotron and inverse-Compton losses have not affected the spectrum of the accelerated electrons, then the flattest spectrum ($\alpha=-0.4\pm0.2$) will represent the spectral index at injection. The flattest possible spectral index that can be generated by the DSA mechanism under these assumptions is $-0.5$ which is consistent to our measurement of the spectral index at the leading edge of the NW relic within the quoted errors. In addition, the flux scale at low frequency is poorly constrained, the flux estimates of low frequency observations have been found to vary significantly based on the adopted flux scale (see e.g \citealt{Scaife2012,Baars1977}). Whilst it is difficult to quantify the impact that this has on our own flux scale we suspect that it could introduce flux errors of 10--20\%, which we are unable to account for. This may be responsible for the reduced $\chi^2$ value derived for our fit to the spectral profile of the NW relic (5.4), which indicates that there are residual errors in our measurements that we have not identified. There are a number of other reasons why we might see such a flat spectral index in the NW relic; stationary shock conditions may not apply in A3667, the acceleration process in A3667 may be more complicated than simple DSA, contaminating point sources projected against the relic emission may be altering the spectral index, the high resolution (6\arcsec), high sensitivity 1.4\,GHz images of A3667 presented in \cite{JohnstonHollitt2003} show 8 such point sources seen in projection through the flattest part of the relic and, as has been previously claimed, there may not be a single electron population present in the NW relic. However, we note that an injection spectral index of $-0.4\pm0.2$ in the NW relic is consistent with the total integrated spectral index of $-0.9\pm0.1$ since it has been shown that the integrated spectral index should $-0.5$ units steeper than the injection spectral index \citep{Miniati2002,Bagchi2002} for a stationary shock model. Given the large number of caveats and possible errors in our spectral index measurements the relic emission in A3667 could be described by a simple DSA model. If DSA is responsible for the NW relic we expect the Mach number to be high; of the order 5. Such high Mach numbers are rare but not impossible \citep{Bonafede2014,Skillman2008}. 

If we select a slice of the leading edge of the NW relic, shown as a green ellipse in the top panel of Fig.~\ref{im:specindex}, we find an average spectral index of $-0.6\pm0.2$ which corresponds to a Mach number of $4.6\pm0.3$. For an average integrated spectral index of $-0.9\pm0.1$ we find a lower Mach number of $2.4\pm0.1$ for both the NW and SE relics. The Mach number at the NW shock front inferred from X-ray observations is $2.4\pm0.8$ over a region that encompasses most of the NW relic \citep{Finoguenov2010}. This is in good agreement with our average value of the Mach number but significantly lower than our leading edge value. Furthermore, these Mach numbers are consistent with a shock generated by a cluster merger \citep{Skillman2008}. 

The sound speed of the ICM towards the NW relic is $710\pm110$\,\kms\ \citep{Finoguenov2010} so the shock has a velocity of $\sim 1400$--$2000$\,\kms. If the shock velocity has been constant over time and we ignore any possible projection effects, we estimate that the NW shock took approximately 1.4\,Gyrs to travel the 2.1\,Mpc from the cluster centre to the current position of the NW relic. Similarly, we derive a time of 1.1\,Gyr for the shock responsible for the SE relic.

For comparison we search the literature for studies of radio relics at low frequency. We present a list of radio relics and the corresponding average spectral indices and available Mach numbers in Table~\ref{tab:comp}. Comparison between the spectral index and Mach number of radio relics is complicated due to the different instruments and frequencies used to observe the relics. We present the reported average spectral indices, the Mach numbers are derived from the flattest spectral indices for the region. We find that the average spectral index for A3667 is flatter than most other known radio relics which may indicate that the relics in A3667 are young. The average Mach number we derive for the relics in A3667 falls within the middle of the range of Mach numbers reported for radio relics. We are unable to reliably draw any further detailed comparisons from the parameters given in Table~\ref{tab:comp}. Ideally a uniform survey with a consistent data collection, reduction and analysis strategy is required to compare the properties of radio relics.

\begin{table*}
\caption[Comparison]{Comparison of the spectral index and Mach number of various radio relics taken from the literature. The Mach number is not presented for relics observed by \cite{Venturi2013} so we derive a lower limit using the average spectral index for the relics and highlight these Mach numbers with a $^*$.}
\begin{center}
  \begin{tabular}{lcccr}
   \hline Cluster & Frequency & Average Spectral & Mach & Reference\\
           Name & Range (MHz) & Index ($\alpha$) & Number ($M$)& \\
 	\hline
A3667 NW	&	120--1400	&	$-0.9\pm0.1$	&	$2.4\pm0.4$	& This work	\\
A3667 SE	&	120--1400	&	$-0.9\pm0.1$	&	$2.4\pm0.4$	& ''	\\
CIZA J2242.8+5301 North	&	153--2272	&	$-1.1\pm0.1$	&	$4.6\pm1.1$	&	\cite{Stroe2013} \\
CIZA J2242.8+5301 South	&	153--2272	&	$-1.3\pm0.2$	&	$2.8\pm0.2$	&	'' \\
A3376 East	&	150--325	&	$-0.9\pm0.1$	&	$3.3\pm0.3$	&	\cite{Kale2012}\\
A3376 West	&	150--325	&	$-1.0\pm0.1$	&	$2.2\pm0.4$	&	''\\
PLCKG287+32.9 NW	&	325--3000	&	$-1.4\pm0.2$	&	$\sim5.4$	& 	\cite{Bonafede2014} \\
PLCKG287+32.9 SE	&	325--3000	&	$-1.3\pm0.1$	&	$\sim3.7$	& 	'' \\
Toothbrush 	&	147--4900	&	$-1.1\pm0.1$	&	$\sim4.6$	&	\cite{VanWeeren2012} \\	
A521	&	240--1400	&	$-1.9\pm0.1$	&	$>1.5*$	&	\cite{Venturi2013} \\
A781	&	325-1400	&	$-1.3\pm0.1$	&	$>1.9*$	&	'' \\
A1300	&	325--843	&	$-0.9\pm0.1$		& $>2.4^*$	&	'' \\	
A2345-1	&	325--1425	&	$-1.5\pm0.1$	&	$2.8\pm0.1$	&	\cite{Bonafede2009} \\
A2345-2	&	325--1425	&	$-1.3\pm0.1$	&	$2.2\pm0.1$	&	''\\
A1240-1	&	325--1425	&	$-1.2\pm0.1$	&	$1.2\pm0.1$	&	''\\
A1240-2	&	325--1425	&	$-1.3\pm0.2$	&	$1.3\pm0.2$	&	''\\
\hline
\end{tabular}
\end{center}
\label{tab:comp}
\end{table*}

\begin{figure*}
\includegraphics[width=0.49\textwidth]{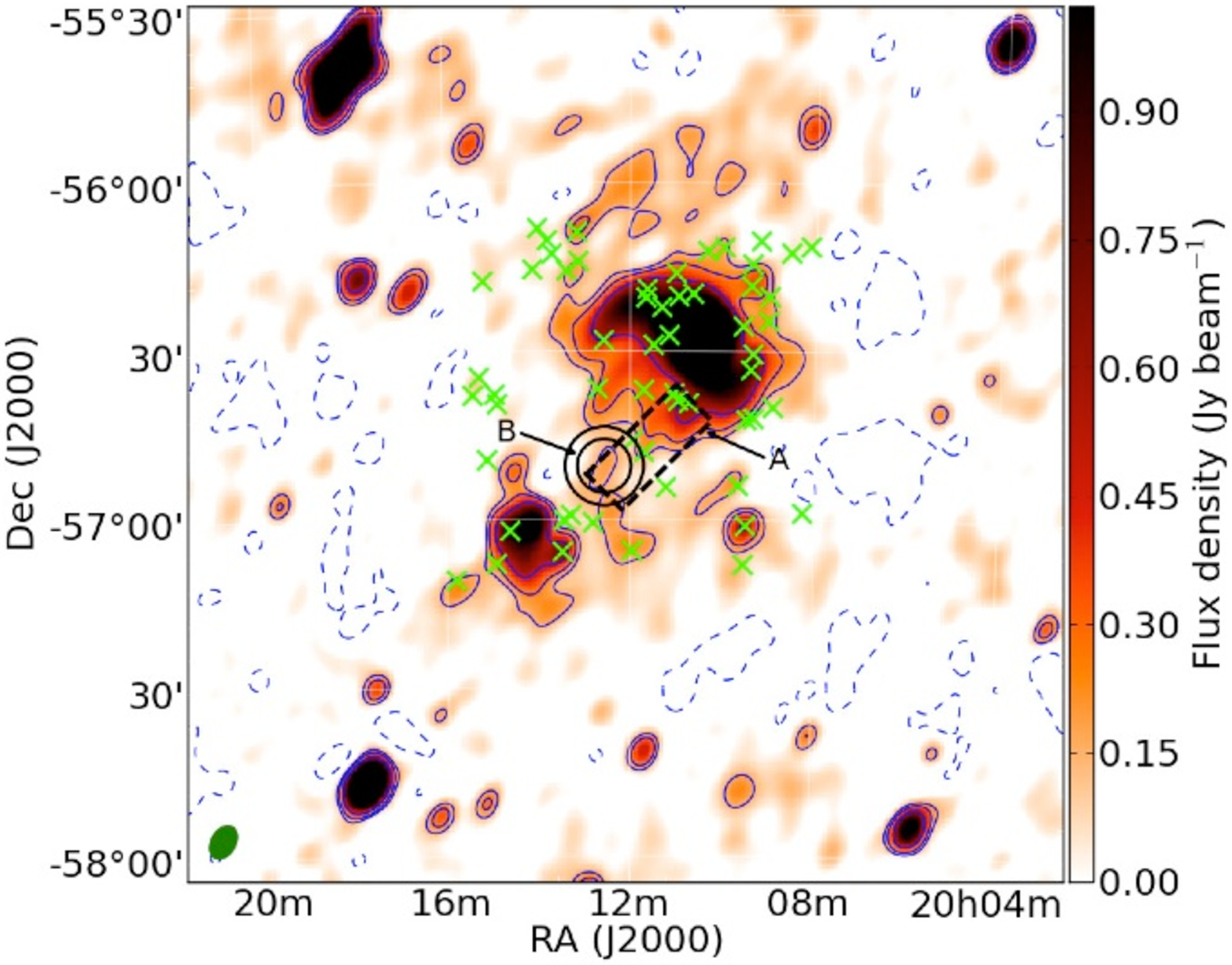} 
\includegraphics[width=0.49\textwidth]{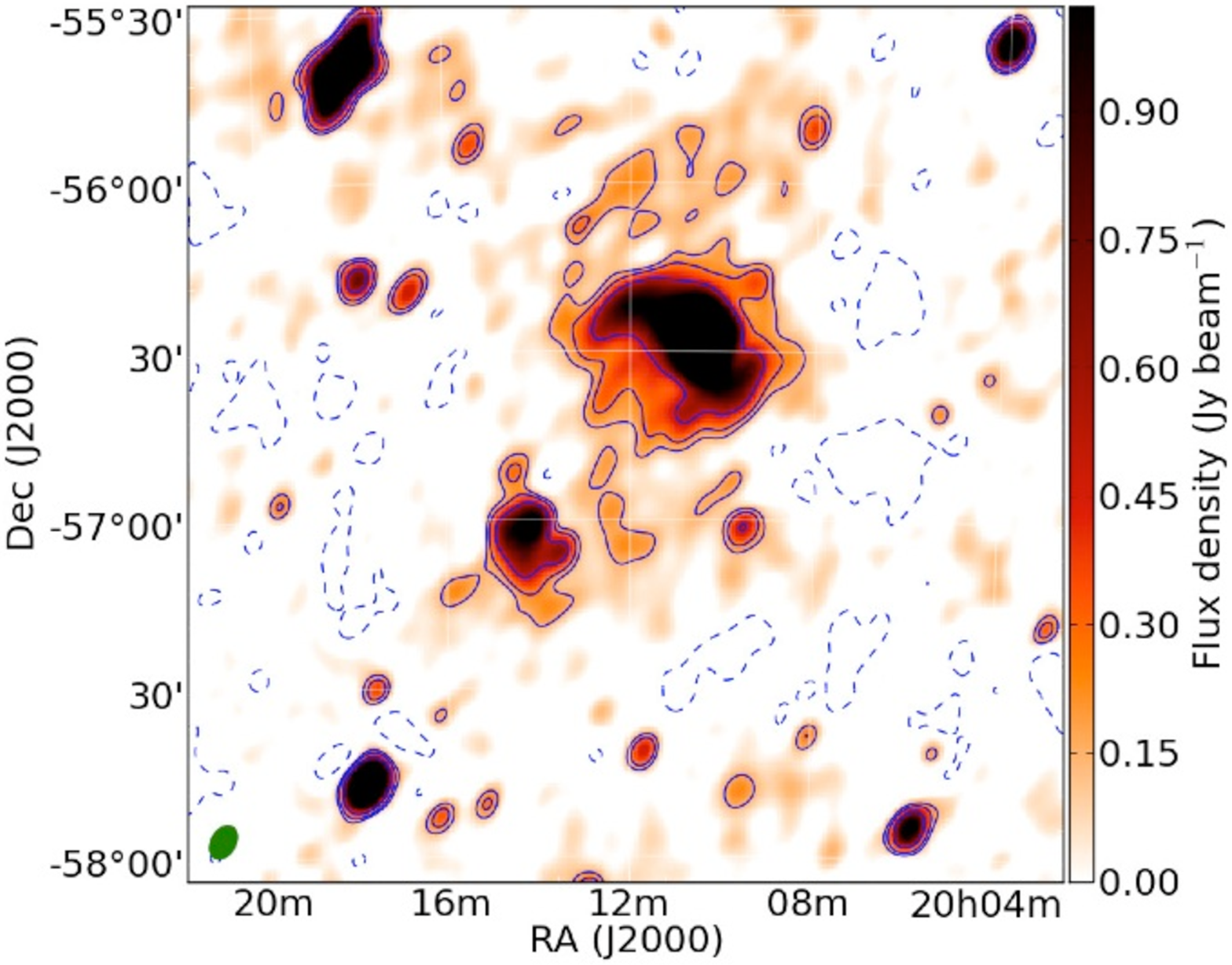} 
\caption{Left panel: resultant 226\,MHz image (robust=0) after peeling MRC\,B2007-568. An annulus, marked B shows the centre of the cluster, defined by the peak in X-ray emission, and the size of the annulus corresponds to the expected radio halo size of 430 and 500\,kpc. Green crosses in the left panel show the location of blended point sources, dashed rectangle, marked A, shows the approximate extent of the proposed bridge reported by \citep{Carretti2013}. Right panel: identical to left panel but without labels. Each image has been smoothed to a resolution of $6\farcm42\times4\farcm37$ to emphasise extended emission. We overlay both images with contours at -1, 3, 5 and 10$\sigma$. The synthesised beam is shown in the bottom left corner of each image as a green ellipse.}
\label{im:peel}
\end{figure*}

\subsection{Radio bridge}\label{SubSect:carretti}

The detection of a radio bridge of unpolarised synchrotron emission connecting the NW relic to the centre of A3667 has been reported by \cite{Carretti2013}. This detection was made at 2.3 and 3.3\,GHz using the Parkes radio telescope with a resolution of $8\farcm9$ and $5\farcm9$, respectively and is supplemented by high resolution ($40\arcsec$) observations at 2.3\,GHz using the Australia Telescope Compact Array. The radio bridge should be prominent at low frequencies but first requires the subtraction of the bright head-tail radio galaxy MRC\,B2007-568. The excellent $u,v$-coverage at short spacing and low-frequency offered by the MWA should allow the detection of this bridge and halo given sufficient integration time.
In order to detect the radio bridge and halo we must first subtract MRC\,B2007-568 from our MWA images. This bright source is well approximated by a point source in our low-resolution MWA images making subtraction fairly simple. We subtract MRC\,B2007-568 from the $u,v$ data used to make the 226\,MHz MWA image, which has the highest sensitivity, by first modelling the source in the image plane using clean components and subtracting the resulting model from the $u,v$ data. We then smooth the peeled 226\,MHz image by convolving it with a Gaussian kernel with FWHM $6\farcm42\times4\farcm37$, to improve the surface brightness sensitivity (Fig.~\ref{im:peel}). This results in a map with a sensitivity of $\sim50$\,mJy\,beam$^{-1}$.

The radio bridge presented by \cite{Carretti2013} is found to have a flux density ranging from 28 to 35\,mJy\,beam$^{-1}$ and integrated flux density of $102\pm8$\,mJy at 3.3\,GHz. The flux density at 2.3\,GHz ranges from approximately $100$ to $115$\,mJy\,beam$^{-1}$ with an integrated flux density of $160$\,mJy. The mean spectral index of the putative radio bridge between 2.3 and 3.3\,GHz is $-1.2\pm0.2$.  By assuming a constant spectral index of $-1.2$ we extrapolate the spectral profile to 226\,MHz and find that the expected integrated flux density of the radio bridge is $\sim2600$\,mJy. If we consider the spectral index reported by \citet{Carretti2013} and assume a 2-sigma lower limit to the spectral index of $–0.9$ we find a lower integrated flux density of $\sim1100$\,mJy.

The putative radio bridge extends for approximately $22\farcm5\times9\farcm0$ (Fig.~\ref{im:peel}; dashed rectangle, left panel), which is encompassed by approximately $7.2$ beams in our smoothed MWA map. Therefore, the expected flux density of the radio bridge in this MWA image, assuming a spectral index of $-1.2$, is $\sim 360$\,mJy\,beam$^{-1}$. We would therefore expect to detect such a radio bridge at a confidence level of $7.2\sigma$. Again if we assume a 2-sigma variation on the spectral index of the bridge of $–0.9$ we find a lower limit to the flux density at 226\,MHz of $150$\,mJy\,beam$^{-1}$. This is equivalent to our $3\sigma$ detection limit.

Inspection of Fig.~\ref{im:peel} reveals the presence of a small region of emission above $5\sigma$ coincident with the putative radio bridge nearest to the NW relic. The emission extends for approximately half of the expected length of the proposed bridge. We are unable to make a significant detection of any further emission associated with the radio bridge towards the centre of the cluster assuming a spectral index of $-1.2$. We note that the emission in this region is coincident with a number of point sources (Fig.~\ref{im:peel}; green crosses) with a combined flux density of $71$\,mJy at 843\,MHz. Assuming a representative spectral index of $-0.7\pm0.2$ these sources would have a combined flux density of 160\,mJy at 226\,MHz, equivalent to $3\sigma$ in our smoothed map. We are unable to peel these sources from the image and suspect they are contributing significantly to the observed emission at the position of the bridge.

\subsection{Radio halo}\label{SubSect:carretti}

The radio halo reported by \cite{Carretti2013} has an integrated flux density at 3.3\,GHz of $44\pm6$\,mJy with a peak brightness of 22\,mJy\,beam$^{-1}$. The integrated flux density of the halo at 2.3\,GHz could not be reliably recovered due to blending with the SE relic. If we assume a representative spectral index for the halo of $-1.3\pm 0.3$, where the error is approximately $3\sigma$ around the mean distribution of known spectral indices, the integrated flux density at 226\,MHz is expected to be $\sim1470$\,mJy. If we assume an upper limit to the spectral index of $-1.0$ we find a lower limit to the integrated flux density at 226\,MHz of $630$\,mJy. The radio halo reported by \cite{Carretti2013} has a size of $<7\arcmin$ giving an upper limit to the halo size of $\sim430$\,kpc. Assuming a lower limit to the halo size of 300\,kpc (angular size $4\farcm8$; B in Fig.~\ref{im:peel}) we expect the emission to be encompassed by $2.6$--5.5 MWA beams at 226\,MHz. We therefore expect a flux density, assuming a spectral index of $-1.3$, of approximately $270$--$570$\,mJy\,beam$^{-1}$ which is a factor of 5.4--11.4 above the sensitivity in our smoothed 226\,MHz image. If we assume an upper limit to the spectral index for the halo of $-1.0$ the expected flux density of the halo is between 110 and 240\,mJy\,beam$^{-1}$. This is a factor of 2.2--4.8 above our sensitivity limit. By considering Fig.~\ref{im:peel} we can see that there is some emission at the $3\sigma$ level associated with the centre of the cluster. This emission is elongated in the same direction as the two relics and appears to sit upon a region of low level extended emission that is below $3\sigma$.

We may independently test this value for the radio halo flux density by applying the radio halo--X-ray scaling relation of \cite{Cassano2013}. By applying this relation, using an X-ray luminosity of $L_{x}=9\times10^{44}$\,ergs\,s$^{-1}$, we find an expected radio halo power of $1.11\times10^{24}$\,W\,Hz$^{-1}$ at 1.4\,GHz which corresponds to a total flux density of 180\,mJy. Assuming a representative spectral index for radio halos of $-1.3$, we find an integrated flux density of 1930\,mJy at 226\,MHz. Again if we assume the halo ranges in size from 430 to 500\,kpc we expect the emission to be encompassed by 2.6--5.5 MWA beams at 226\,MHz. This corresponds to a flux density of $\sim 350$--740\,mJy\,beam$^{-1}$. Given the sensitivity in our peeled and smoothed 226\,MHz images we would expect the radio halo to be detected between $7$--$15\sigma$, which is clearly not the case. In the case of a spectral index of $-1.0$ we would expect a flux density of 1060\,mJy leading to a flux density of 190--410\,mJy\,beam$^{-1}$ a factor of $3.8$--$8.2$ above our sensitivity limit.

\subsection{Radio halo and bridge: discussion}

\begin{figure*}
\includegraphics[width=0.49\textwidth]{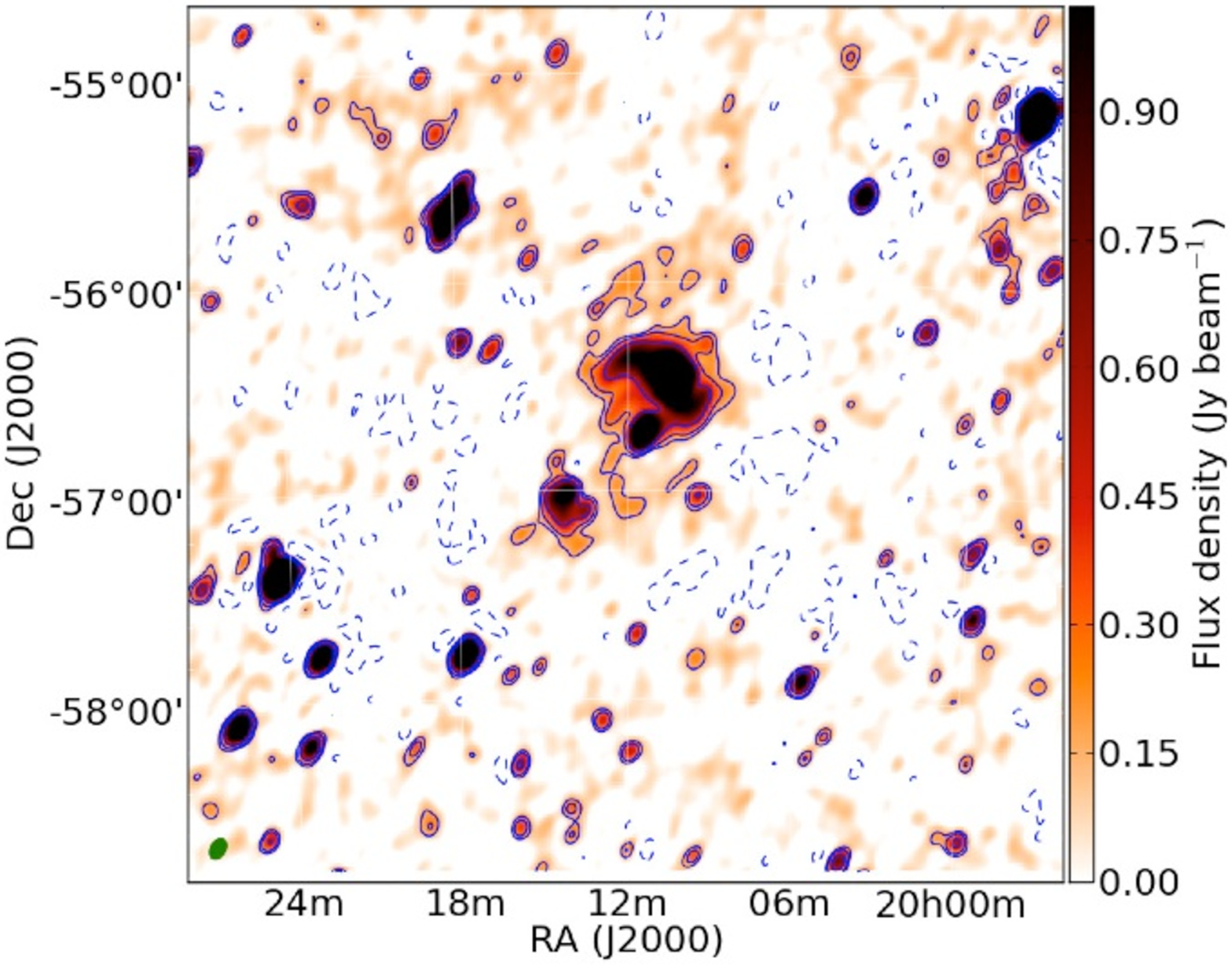} 
\includegraphics[width=0.49\textwidth]{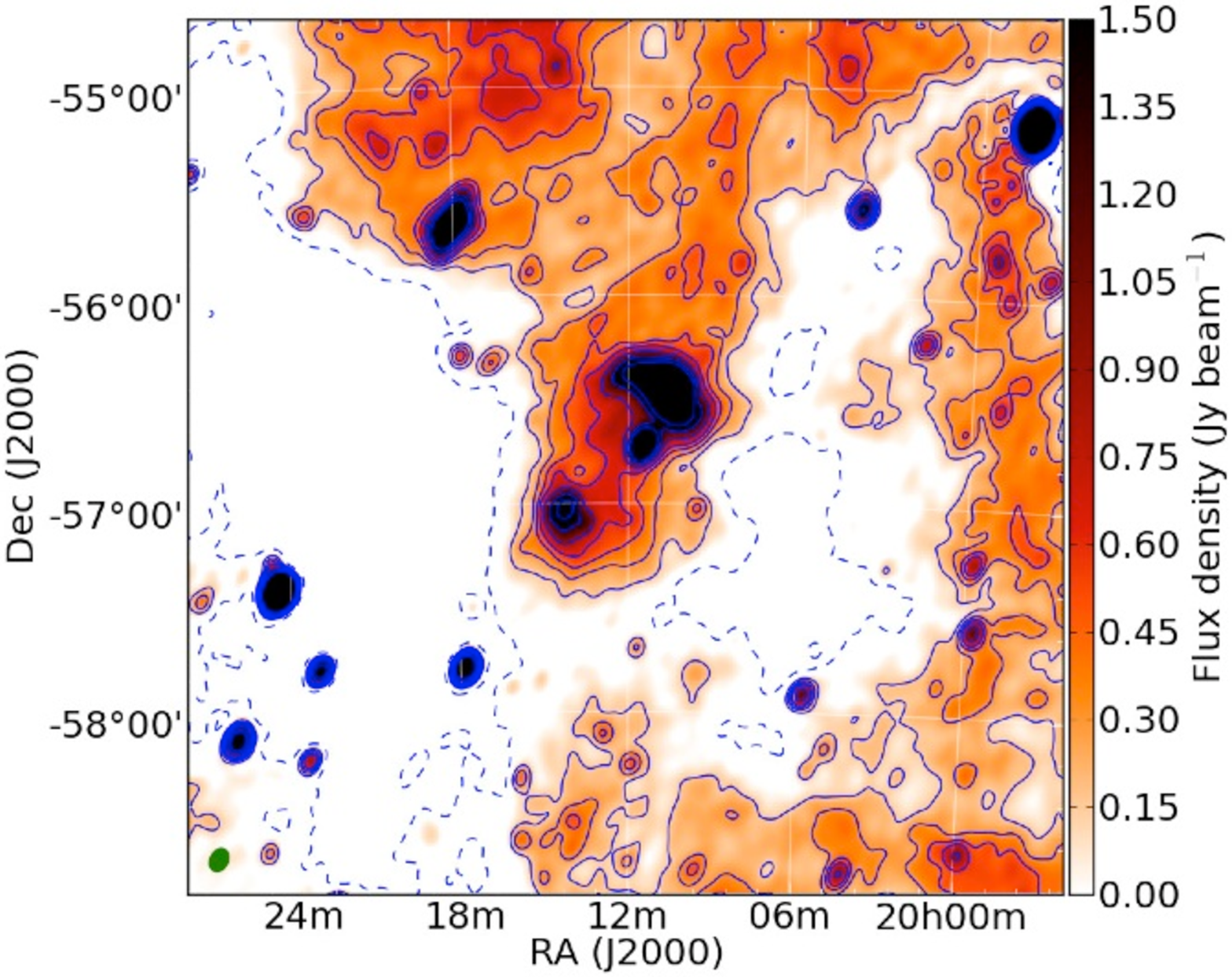} 
\caption{Left panel: image of A3667 at 226\,MHz using a robust weighting of 0. Right panel: a naturally weighted (robust=2) image of A3667 at 226\,MHz. Each image has been smoothed to a resolution of $6\farcm42\times4\farcm37$ to emphasise extended emission. We overlay the left hand image with contours at -3, 3, 5 and 10$\sigma$. The naturally weighted right hand image is overlaid with contours at -3, 3, 4, 5, 6, 9 and 12$\sigma$. We show the synthesised beam in the bottom left corner of each image as a green ellipse.}
\label{im:weight}
\end{figure*}

The robust=0 maps in Fig.~\ref{im:peel} show that A3667 sits on a ridge of extended emission with negative bowls either side which is indicative of poorly sampled large-scale emission. This leads us to suspect that the source of the low-level emission coincident with the reported bridge and halo may be the result of peaks in large-scale emission that are poorly sampled in our robust=0 weighted MWA image. For this reason we produce a naturally weighted (robust=2) image (Fig.~\ref{im:weight}; right panel) which is sensitive to emission on very large scales. This image reveal that A3667 is projected against a region of extended emission that pervades the entire field. The flux density of this large-scale extended emission ranges from 100 to 300\,mJy/beam$^{-1}$. The origin of this emission is the Milky Way, which is approximately $30\degr$ from the field centre. This emission is accounted for by \cite{Carretti2013} who estimate a contribution from Galactic emission of 8 and 2\,mJy at 2.3 and 3.3\,GHz, respectively. Inspection of the publicly available continuum HI Parkes All Sky Survey (CHIPASS; \citealt{Calabretta2014}) at 1.4\,GHz reveals that A3667 is associated with significant Galactic emission. It is difficult to disentangle this Galactic emission from the emission associated with a putative radio bridge and halo. It is possible that the proposed radio bridge and halo in A3667 are the result of compact peaks in this large-scale extended emission. The $3\sigma$ contours from our peeled robust=0 map clearly trace the boundaries of the bright extended emission in the naturally weighted map.

The expected flux densities for the radio bridge and halo are a factor of $3.0$--$7.4$ and $2.2$--$14.8$ above the sensitivity limit in our smoothed 226\,MHz image, respectively. This range is due to the assumption of representative values for the spectral index and size of the features. While it may be argued that these observations may have detected emission associated with a radio bridge, we are unable to verify the bridge in A3667 due to the numerous point sources and poorly sampled extended Galactic emission. We find evidence of a possible radio halo at our $3\sigma$ detection limit within the centre of the cluster but again this may be a false detection caused by Galactic emission. It is possible that both the radio halo and bridge have flat spectral indices and the sources are on the larger side of the normal range of known sizes leading to lower integrated flux densities at 226\,MHz. This results in the sources being on the border of our detection limit at the $3\sigma$ level. Alternatively, the radio bridge and halo may be the result of a false detection, caused by a combination of unrelated extended Galactic radio emission, blended point sources and possibly residual emission from peeling. At this time we are unable to distinguish which of these is true. We require a data set with improved sensitivity and resolution and a more thorough peeling method to account for blended point sources. 

In summary, we detect emission that could be associated with the putative radio halo and bridge within A3667. However, the flux of these features is lower than expected and our images suffer from the presence of contaminating point sources and significant levels of poorly sampled Galactic emission. This prevents us from making a clear detection of either a radio halo or bridge.

\subsection{A3667: a merging cluster with no radio halo}
Radio halos are frequently found to be associated with dynamically disturbed merging galaxy clusters. It is thought that the energy dissipated during a cluster merger may be responsible for powering radio halos via the turbulent re-acceleration mechanism \citep{Cassano2010A}. We would expect A3667, which has undergone a recent merger, to be host to a radio halo. Optical and X-ray observations of A3667 provide clear evidence that the cluster is undergoing a merger event and the cluster members are in a highly dynamical state. A plausible model has been produced which suggests that the merger event in A3667 occurred within the last $\sim 1$\,Gyr \citep{Roettiger1999}. Statistical analysis of galaxy clusters \citep{Brunetti2009} suggests that radio halos are generated over fairly short time-scales after a merger event, on the order of a few 100\,Myr, and fade rapidly on longer time-scales of Gyrs. Given such a time-scale if turbulent re-acceleration is responsible for the generation of radio halo in A3667 it is possible that the radio halo emission in A3667 has faded below our detection limit. It is also possible that some mechanism is reducing the efficiency of electron acceleration in A3667. This would lead to a steep spectrum radio halo which is not visible at higher frequencies (1.4\,GHz). The low frequency of these MWA observations should enable the detection of a steep spectrum radio halo \citep{Cassano2010B}. The integrated flux of a radio halo at our 3-sigma detection limit is 390--825\,mJy; combining this with the 44\,mJy measurement at 3.3\,GHz, we calculate a spectral index for the radio halo between $-1.1$ and $-0.8$. A3667 is clearly not host to an ultra-steep spectrum radio halo. This indicates that merging maybe a necessary but not sufficient condition for halo generation \citep{Ensslin1998,Ensslin2001,JohnstonHollitt2003}.

\section{Conclusions}\label{Sect:summary}

\subsection{Future prospects}

The GaLactic and Extragalactic MWA (GLEAM) survey currently being carried out with the MWA is mapping the entire southern sky below a declination of $+25\degr$. These observations consist of approximately an hour of integration per pointing. It is estimated that the sensitivity of these observations will reach the confusion limit of 5--60\,mJy\,beam$^{-1}$ (depending on frequency), and with the additional $u,v$-coverage we can expect even better surface brightness sensitivity. The GLEAM survey will allow us to explore the low-frequency radio environment of already detected radio halos and relics in the southern sky. We will also be able to search for a new sample of radio halos and relics using new detection algorithms such as the circle Hough transform \citep{JohnstonHollitt2012}, Latent Dirichelet Allocation \citep{Friedlander2014,Frean2014} and other diffuse source detection methods such as localised kernel transforms \citep{starck2003,Finoguenov2006} or possibly appropriately filtered segmentation methods \citep{whiting2012,Hancock2012} and spatial scale filtering \citep{Rudnick2009}. For further details and discussion of the limitations of these algorithms for diffuse source detection, see Section 4.5.2 of \citet{Norris2013}.

The principal limitation of the MWA with regards to the study of radio halos and relics is the low angular resolution of the observations. As we encountered here, point sources detected at higher frequencies are often blended with extended emission. These contaminating sources can lead to errors in flux density and spectral index estimates. To accurately subtract these sources, we will require high-resolution follow-up observations at a similar frequency. The Giant Metrewave Radio Telescope (GMRT) operates between 150 and 1420\,MHz with a synthesised beam (FWHM) ranging from 20 to $2\arcsec$ (depending on frequency) making it an ideal instrument with which to obtain such follow-up observations. The TIFR GMRT Sky Survey\footnote{http://tgss.ncra.tifr.res.in/} (TGSS) is currently being performed with the GMRT and is surveying the sky North of $-55\degr$ at 150\,MHz with a sensitivity of 5--7\,mJy\,beam$^{-1}$ at a resolution of $20\arcsec$. For sources South of $-55\degr$ accurate removal of point sources will be more difficult. Nevertheless the number of detected sources between $+25\degr$ and $-55\degr$ should prove valuable for future studies.

\subsection{Summary}
We have presented the first MWA observations of the extended low-frequency radio emission associated with the galaxy cluster A3667. These observations were carried out at a central frequency of 120, 149, 180 and 226\,MHz with a 30\,MHz bandwidth. We clearly detect the two radio relics on the NW and SE periphery of A3667. The integrated flux density of the NW and SE relics are $28.1\pm1.7$ and $2.4\pm0.1$\,Jy at 149\,MHz, respectively. We determine the spectral index for the NW and SE relics in the range 86--1400 \,MHz and 120--1400\,MHz, respectively, and find a value of $-0.9\pm0.1$ for both relics. We determine a Mach number using the averaged spectral index for both relics of $2.4\pm0.4$. By using the spectral index determined from a slice of the NW relics leading edge we find a higher Mach number of $4.6\pm2.4$. These results are consistent with the X-ray derived Mach number and provide evidence that the origin of the relics in A3667 is due to a merger. We are unable to confirm the presence of a radio bridge reported by \citep{Carretti2013} or the putative radio halo alluded to by \cite{JohnstonHollitt2003,JohnstonHollitt2004} and reported by \cite{Carretti2013}. At 226\,MHz we find the expected integrated flux density of the radio bridge and halo to be $\sim 2600$ and $\sim 1470$\,mJy, respectively. We also predict the integrated radio halo flux density to be $\sim 1930$\,mJy using the X-ray luminosity and radio power relation from \cite{Cassano2013}. Given the sensitivity in our smoothed 226\,MHz is 50\,mJy\,beam$^{-1}$ and a range of possible parameters for the expected spectral indices and size we would expect to be able to detect these features in our map. Future observations with the MWA, principally the GLEAM survey, should be able to detect a radio bridge and halo in A3667 if present.

\section{Acknowledgements}
We thank the anonymous referee for their very useful comments. MJ-H acknowledges support from the Marsden Fund. This scientific work makes use of the Murchison Radio-astronomy Observatory, operated by CSIRO. We acknowledge the Wajarri Yamatji people as the traditional owners of the Observatory site. Support for the MWA comes from the U.S. National Science Foundation (grants AST-0457585, PHY-0835713, CAREER-0847753, and AST-0908884), the Australian Research Council (LIEF grants LE0775621 and LE0882938), the U.S. Air Force Office of Scientific Research (grant FA9550-0510247), and the Centre for All-sky Astrophysics (an Australian Research Council Centre of Excellence funded by grant CE110001020). Support is also provided by the Smithsonian Astrophysical Observatory, the MIT School of Science, the Raman Research Institute, the Australian National University, and the Victoria University of Wellington (via grant MED-E1799 from the New Zealand Ministry of Economic Development and an IBM Shared University Research Grant). The Australian Federal government provides additional support via the Commonwealth Scientific and Industrial Research Organisation (CSIRO), National Collaborative Research Infrastructure Strategy, Education Investment Fund, and the Australia India Strategic Research Fund, and Astronomy Australia Limited, under contract to Curtin University. We acknowledge the iVEC Petabyte Data Store, the Initiative in Innovative Computing and the CUDA Center for Excellence sponsored by NVIDIA at Harvard University, and the International Centre for Radio Astronomy Research (ICRAR), a Joint Venture of Curtin University and The University of Western Australia, funded by the Western Australian State government.

\appendix
\setcounter{figure}{0} \renewcommand{\thefigure}{A.\arabic{figure}} 
\setcounter{table}{0} \renewcommand{\thetable}{A.\arabic{table}} 
\section*{Appendix A}
Flux density calibration at the low declination ($< -50\degr$) and frequency ($<226$\,MHz) of MWA observations is non-trivial for a number of reasons. Firstly, there is a lack of low-frequency survey data in the southern sky. The VLA Low-Frequency Sky Survey (VLSS; \citealt{Cohen2007,Lane2014}) at 74\,MHz only covers the sky North of $-30\degr$, whilst the Culgoora catalog \citep{Slee1995} at 80 and 160\,MHz only covers a declination $>-50\degr$. Higher frequency surveys such as the Molonglo Reference Catalog (MRC; \citealt{Large1991}) at 408\,MHz and the Sydney University Molonglo Sky Survey at 843\,MHz (SUMSS; \citealt{Bock1999}) extend from $-85\degr$ to $+18\degr .5$ and $< -30\degr$, respectively. However, the frequencies of these surveys are not ideal when trying to constrain the flux density scale of MWA observations due to the inherent uncertainty when extrapolating the spectral profile and also because of the possible spectral turn over of sources at low frequency due to synchrotron self absorption. Secondly, commissioning of the MWA revealed that the analytic primary beam model described by \cite{Williams2012}, which is currently used for primary beam correction, is not consistent with the real beam response \citep{Hurley2014}. The analytic beam model has been used as an approximation of the true beam response in the absence of a more complex model to enable progress during the commissioning period. The shape of the beam response of the MWA depends on the beam former settings, frequency, polarisation and electronic coupling between the individual dipole antennas. The contribution of these factors to the beam shape is non-linear and the analytic model currently used to determine the beam shape does not take into account the mutual coupling between dipole elements and tile-to-tile differences. This issue is amplified by the large delays used to generate beams at low declinations, the MFS nature of the data reduction process and the unknown number of non-functional dipoles in the array, which was not a metric that was recorded during the commissioning period. Further analysis is being carried out to determine the beam shape \citep{Sutinjo2014} but at this time we must attempt to compensate for the unknown flux density errors introduced by the current primary beam model. 

\begin{figure*}
\includegraphics[width=0.49\textwidth]{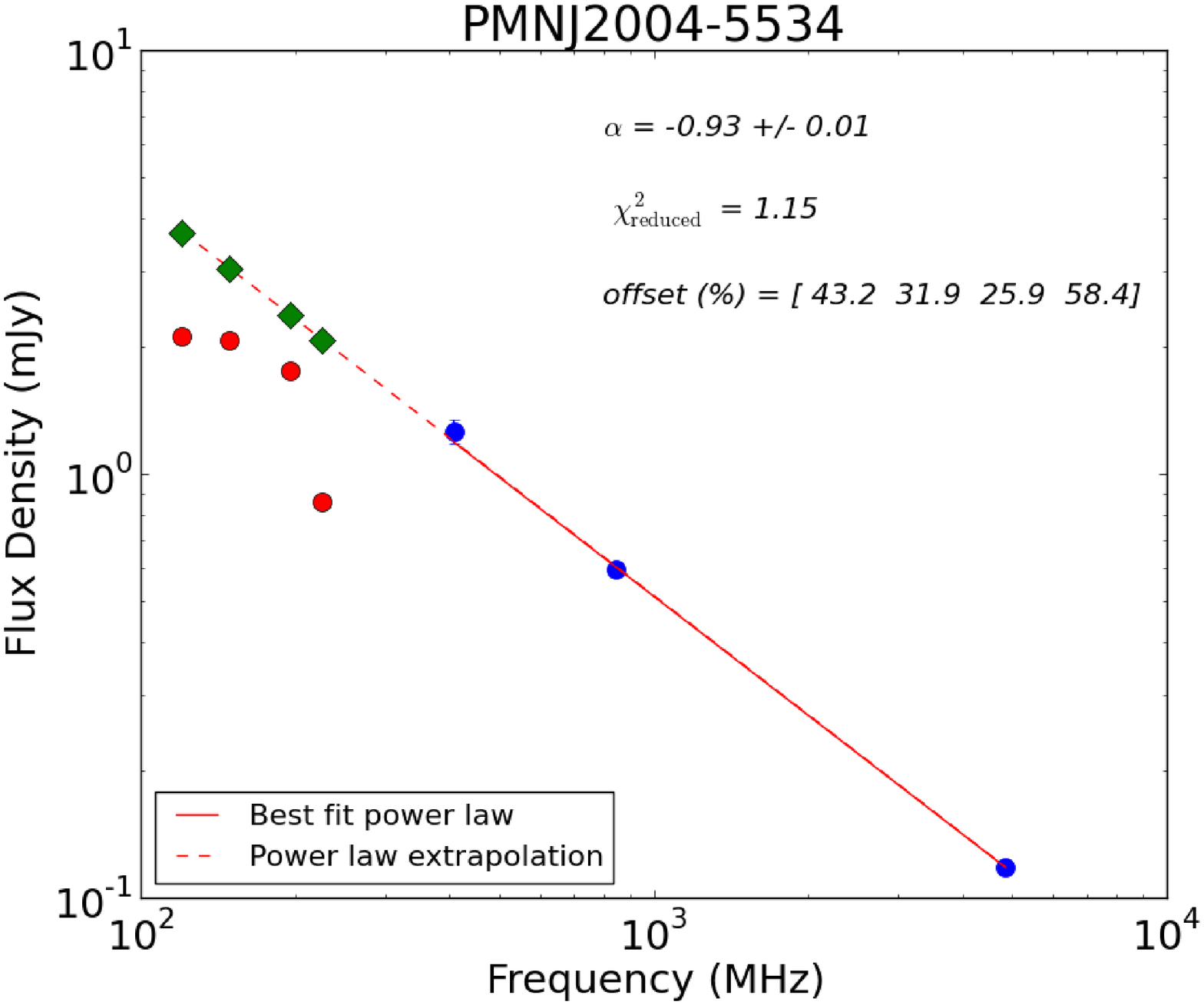} 
\includegraphics[width=0.49\textwidth]{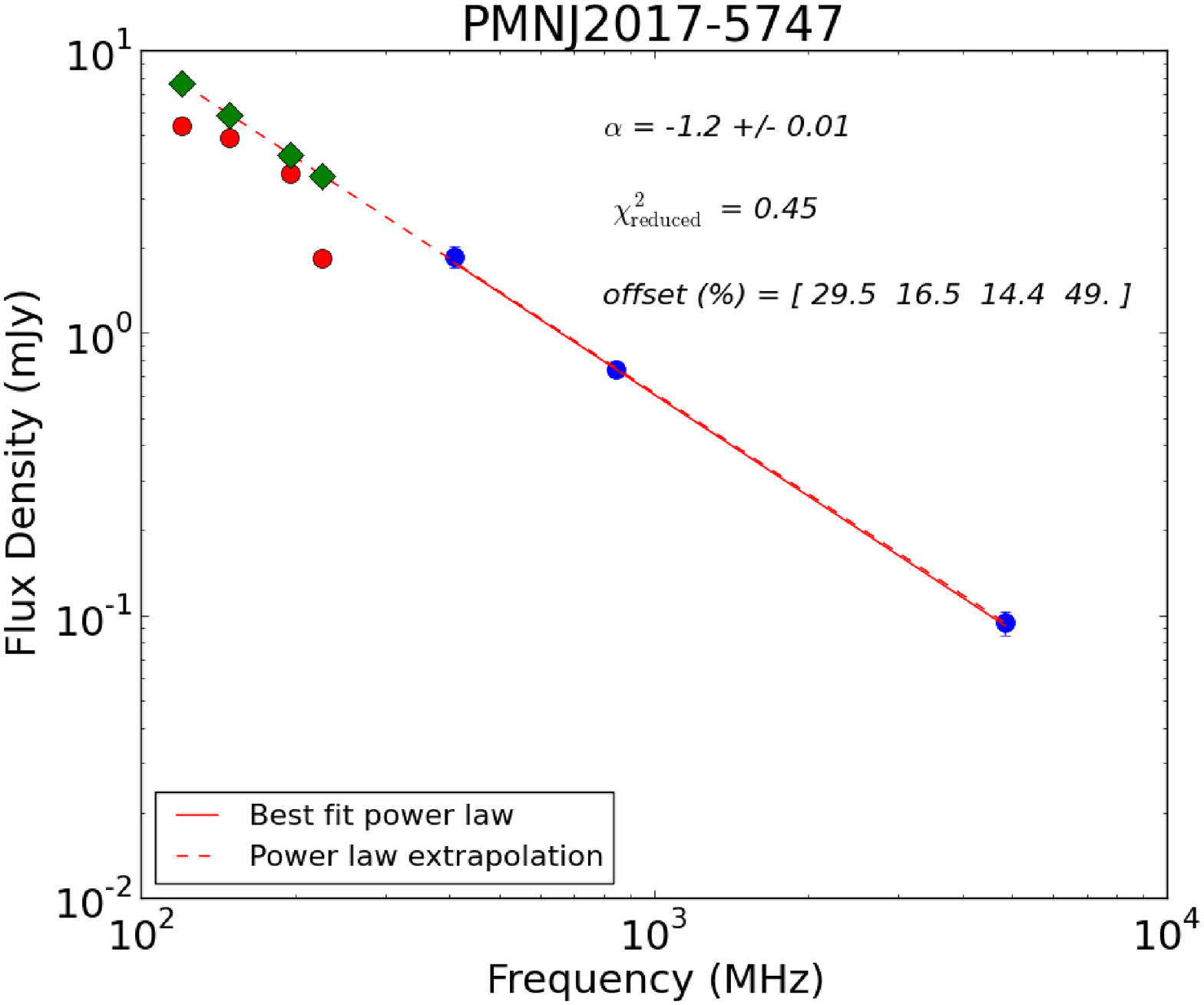} 
\caption{The inherent offset in the MWA flux density scale using two sources. Blue points show the flux density with errors at 408, 843 and 4850\,MHz for the two test sources PMN\,J2004-5534 (left) and PMN\,J2017-5747 (right). The red line through these points shows the best-fit power-law. The red dashed line shows the extrapolated fit assuming a constant spectral index. Red circles show the uncorrected MWA flux densities. Green diamonds show the expected value of the MWA fluxes. The percentage offset between the red and green points at 120, 149, 180 and 226\,MHz is shown.}
\label{im:nocorrection}
\end{figure*}

At the time of writing the best method available is to bootstrap the flux density of the MWA images using known sources in the field \citep{Hurley2014,mckinley2013,Williams2012} . To demonstrate the uncorrected MWA flux density scale we present the MWA flux densities for two sources (PMN\,J2004-5534 and PMN\,J2017-5747) in Fig.~\ref{im:nocorrection}. These test sources do not have flux densities available below 408\,MHz so we must assume that they have a constant spectral index extrapolated from the flux density at 408 \citep{Large1991}, 843 \citep{Bock1999} and 4850\,MHz \citep{Wright1994}. We are able to define a percentage offset value which, indicates how far our MWA flux densities are from the fitted spectral profile. For our uncorrected MWA images in Fig.~\ref{im:nocorrection} we find that the percentage offset is different for each source ranging from 14.4 to 58.4\%.

To correct the MWA flux density scale we first compile a list of suitable calibration sources using the Molonglo Southern 4\,Jy Sample (MS4; \citealt{Burgess2006}). The MS4 catalogue contains 228 bright sources with flux densities $>4$\,Jy at 408\,MHz between $-85\degr<\delta<-30\degr$. \cite{Burgess2006} have performed cross-matching with various archival and published data to calculate the flux density at intervals between 86 and 5000\,MHz. The flux density error estimate given by \cite{Burgess2006} for bright and compact sources is $\sim10\%$ and this is factored into our method. To minimise errors in the flux density calibration we apply the following selection criteria to MS4 sources:

\begin{itemize}
\item The source must be unresolved in the MS4 catalogue and in all of the MWA frequency bands so we may use the peak brightness.
\item There must be an 86\,MHz point for the MS4 source so that we are able to constrain the fitted spectral profile below the MWA frequencies.
\item The spectral profile derived from the MS4 flux density must be fitted by a simple power-law to remove uncertainty due to curvature or spectral turn over. 
\item The fit to the spectral profile must have a reduced $\chi^2$ value less than 2.0 indicating a good fit.
\end{itemize}

\noindent After applying these selection criteria we identify five MS4 sources that are suitable flux density calibrators within $12\degr$ of A3667.

Before proceeding further we mention two issues that affect the flux density estimates in the MWA images. Firstly, stacking snapshots in the image plane, particularly at these low frequencies, may lead to a blurring of sources due to position shifts caused by ionospheric distortions on short time-scales. Secondly, the repeated stacking of snapshots and the corresponding different restoring beams in the image plane results in an enlargement and circularisation of the final point
spread function that does not represent the true shape of the synthesised beam. The combined error introduced by these effects is small; the estimated correction is of the order a few percent for MWA observations that comprised almost an hour of rotation synthesis and ionospheric distortion (see, \citealt{Hurley2014}). For the comparatively short time-scale observations and small number of snapshots presented here, we expect the error to be negligible, within 1--2\%, and factor this into our error analysis.

\begin{figure*}
\includegraphics[width=0.49\textwidth]{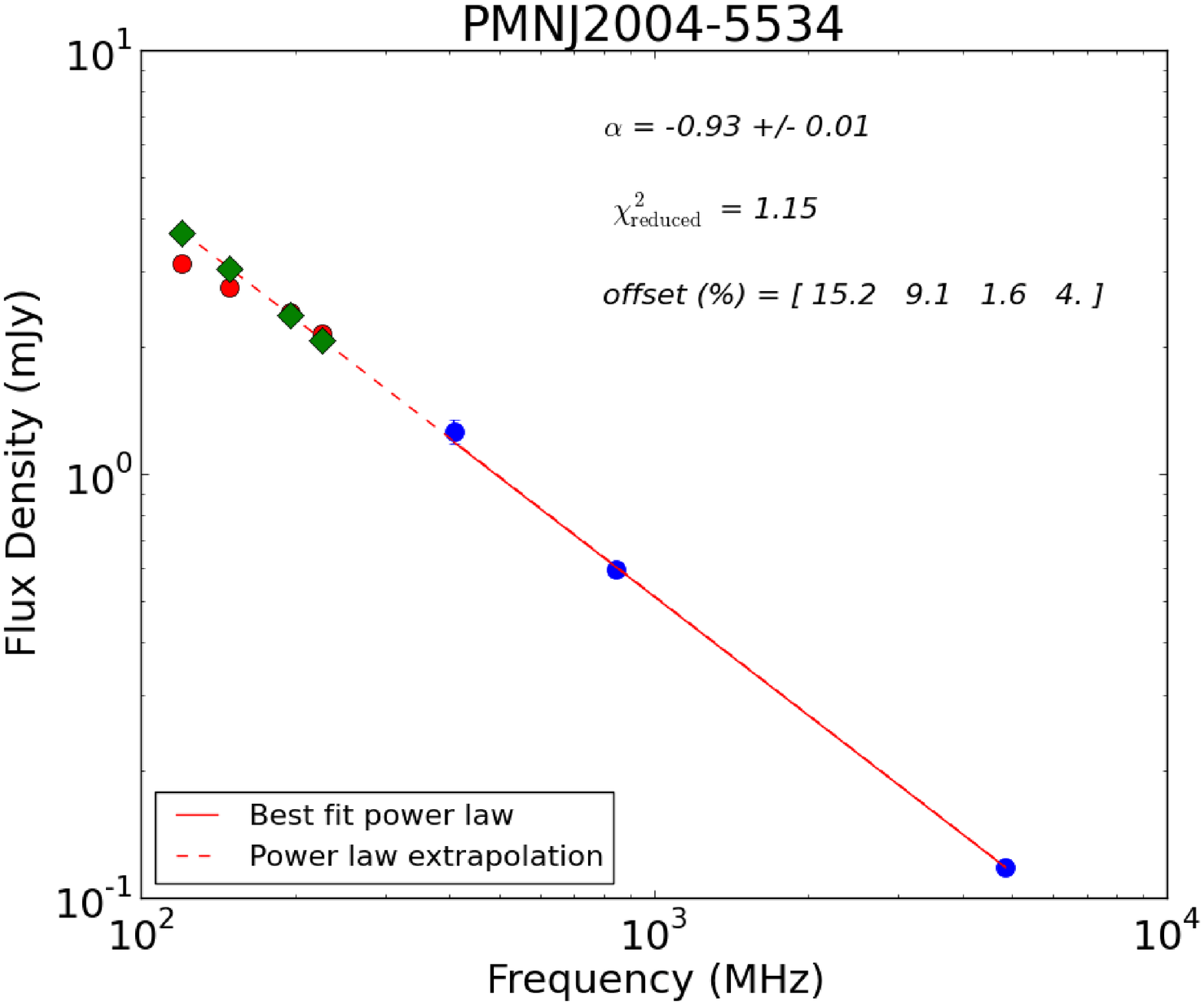} 
\includegraphics[width=0.49\textwidth]{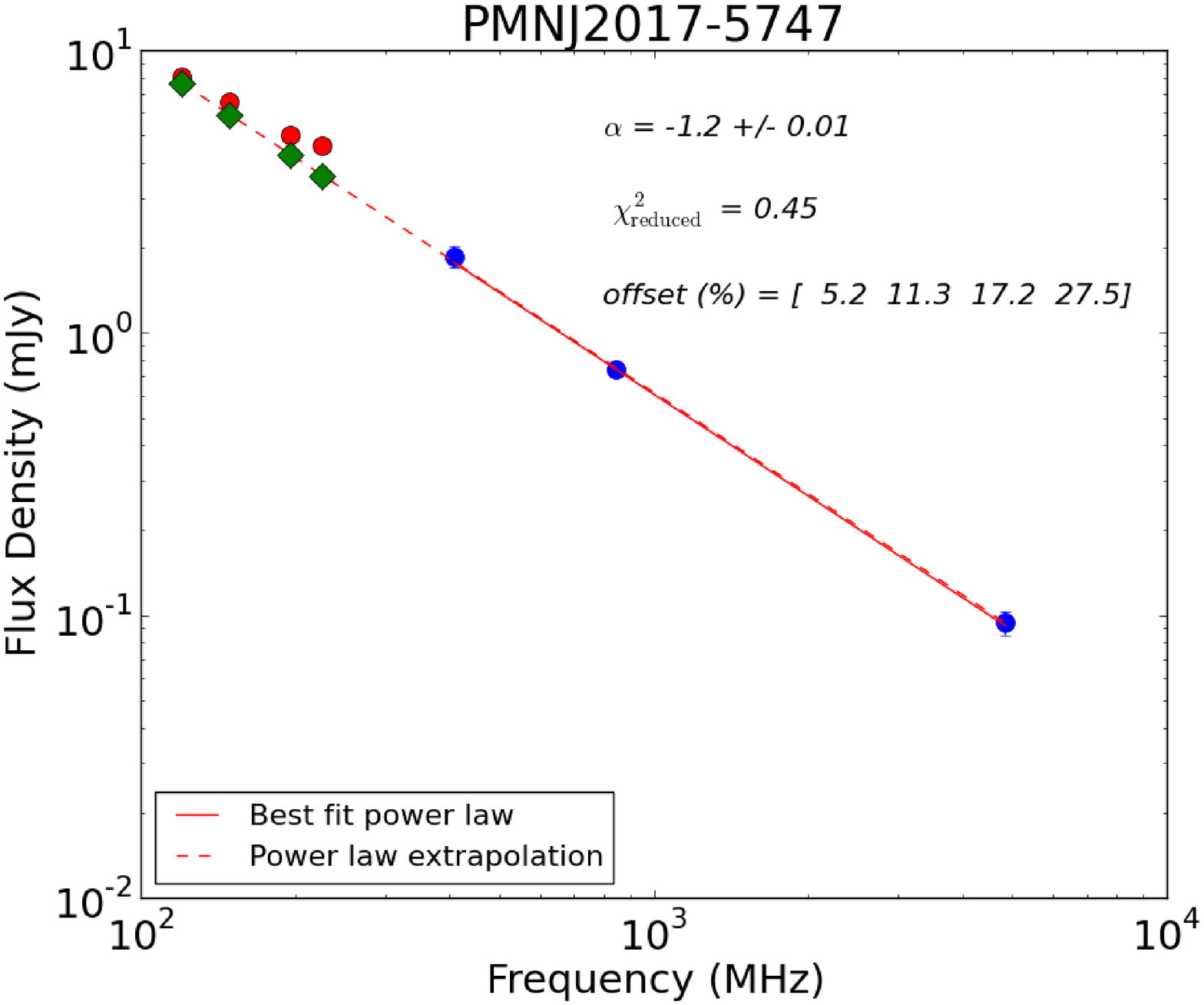} 
\caption{Results of a simple flux density scaling approach. Here we bootstrap the flux density scale using the nearby source MRC\,B2020-575. Blue points show the flux density with errors at 408, 843 and 4850\,MHz for the two test sources PMN\,J2004-5534 (left) and PMN\,J2017-5747 (right). The red line through these points shows the best-fit power-law. The red dashed line shows the extrapolated fit assuming a constant spectral index. Red circles show the MWA flux densities after scaling the flux density by bootstrapping the flux density scale using the MS4 source MRC\,B2020-575. The residual percentage offset values is given at 120, 149, 180 and 226\,MHz.}
\label{im:testsimple}
\end{figure*}

Initially we applied a simple global flux density correction to each MWA frequency image by fitting a power-law to the closest MS4 source, MRC\,B2020-575, at 86, 178, 408, 843 and 1400\,MHz and scaled the MWA flux densities to the fitted line. To test the accuracy of this approach we compare the corrected MWA flux densities using this simple method to the spectral profile of the two nearby point sources PMN\,J2004-5534 and PMN\,J2017-5747. The resulting MWA flux densities after applying this simple correction method can be seen in Fig.~\ref{im:testsimple}. This approach does improve the accuracy of the flux density estimates but results in a large offset in the flux density of the two test sources with a residual offset of up to 30\% at the highest frequency (Fig.~\ref{im:testsimple}). Clearly the flux density is varying over the area of a few degrees and this simple bootstrapping method cannot correct for this variation. 

\begin{figure}
\includegraphics[width=0.49\textwidth]{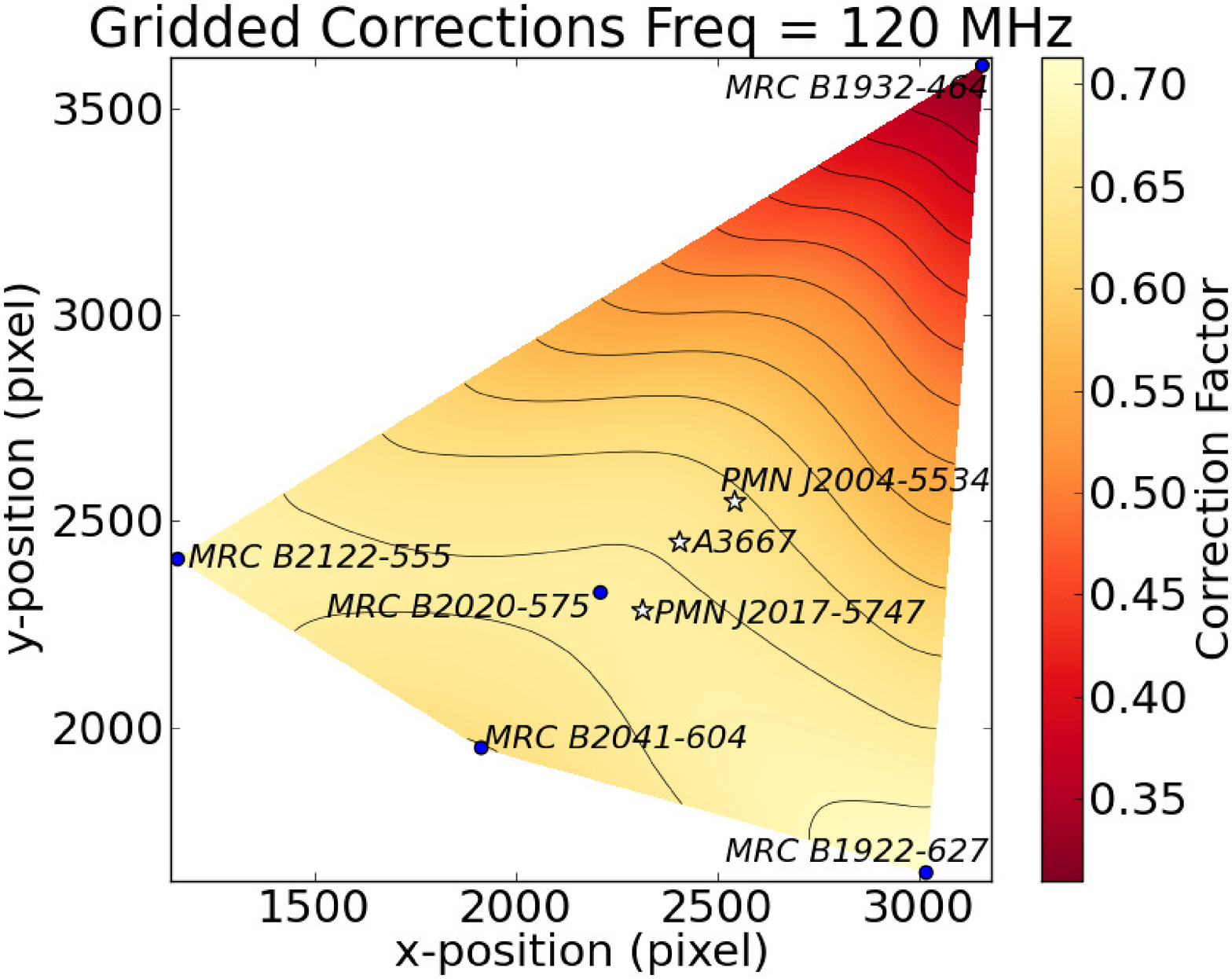} 
\caption{The interpolated correction values are shown for the 120\,MHz band. MS4 sources used for flux density calibration are labelled. Black dots indicated sources used in the interpolation routine, white stars show the position of A3667 and the two test sources PMN\,J2004-5534 and PMN\,J2017-5747. The shape of the corrected region is defined by the position of the outermost available flux density calibrators. The colour scale represents the values of the correction values which are unitless. }
\label{im:correction}
\end{figure}

\begin{figure*}
\includegraphics[width=0.49\textwidth]{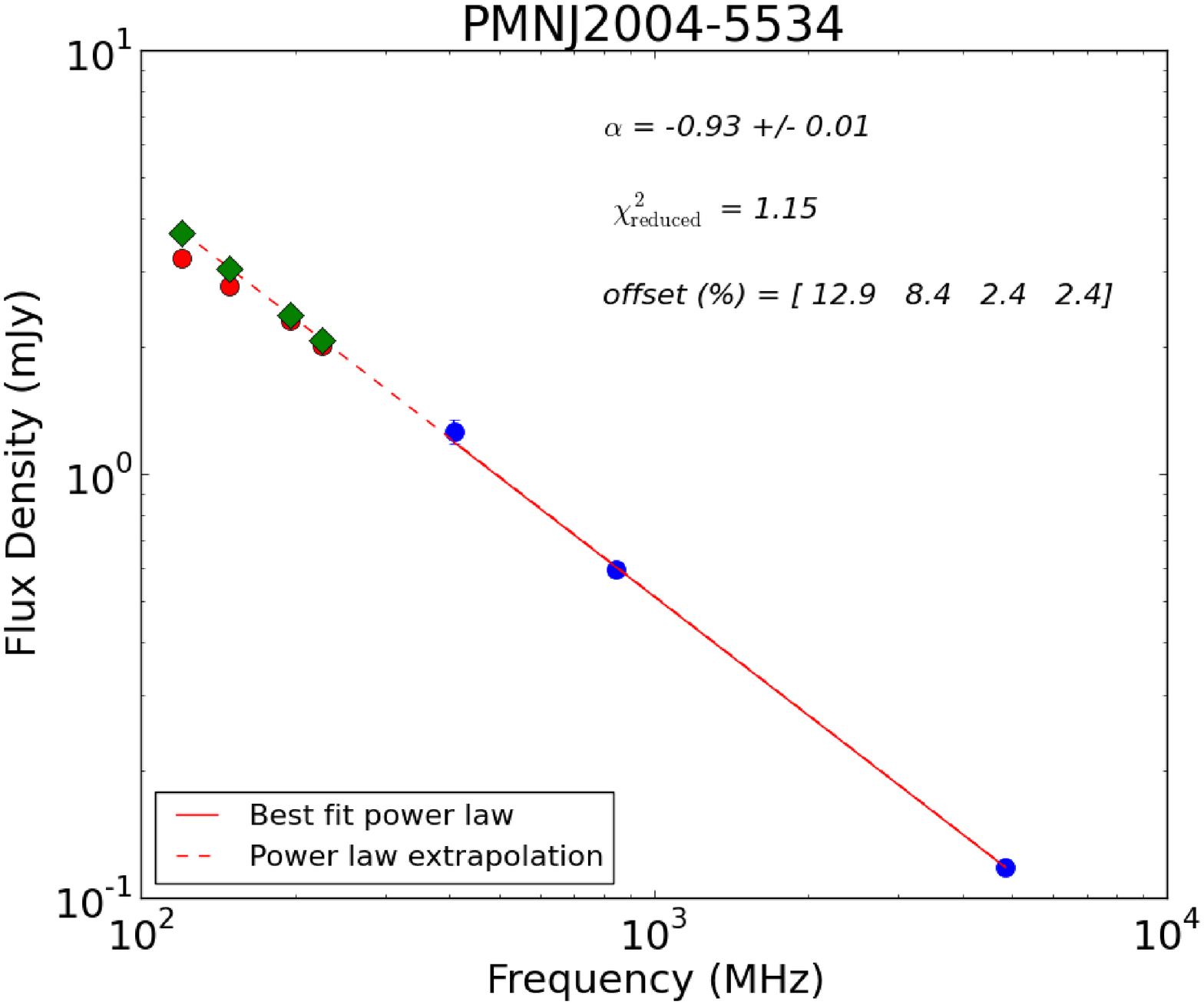} 
\includegraphics[width=0.49\textwidth]{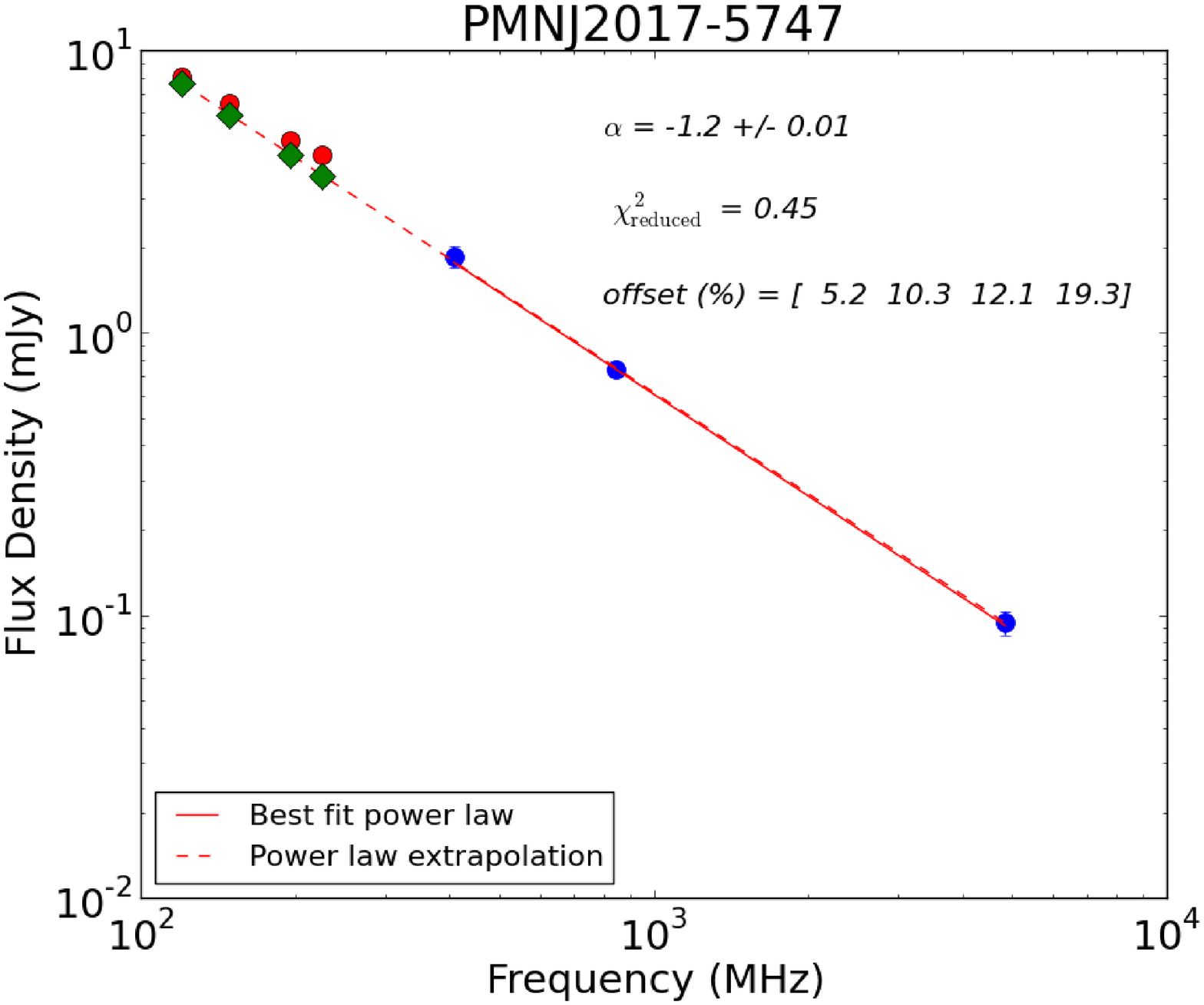} 
\caption{Improved MWA flux density scale demonstrated for two test sources after application of the cubic interpolation method. Blue points show the flux density at 408, 843 and 4850\,MHz for the two test sources  PMN\,J2004-5534 and PMN\,J2017-5747. The solid red line through these points shows the best-fit power-law. The dashed red line shows where we have extrapolated the flux density assuming a constant spectral index. Red circles show the MWA flux densities after applying the interpolation method. The residual percentage offset of the MWA flux densities to the extrapolated fit is presented in the top right at 120, 149, 180 and 226\,MHz.}
\label{im:corrected}
\end{figure*}

To improve the flux density scale, we have determined the offsets for each of the five MS4 sources that meet our selection criteria and then applied a two-dimensional cubic interpolation algorithm to estimate the position-dependent flux density scaling at each frequency. This method results in a two-dimensional grid of correction values for each pixel within the boundary defined by the position of the MS4 sources. Cubic interpolation was found to produce the lowest residual offsets in the flux density scaling compared to other interpolation algorithms such as linear and triangular. To minimise the error induced by the changing primary beam shape and variable ionosphere in each snapshot and by stacking effects, we apply our flux density calibration method to each primary beam corrected snapshot individually, then combine them to obtain a final flux density scaled image. In Fig.~\ref{im:correction}) we show the locations of the two test sources, flux density calibration sources and the resulting grid of offsets values for a single snapshot at 120\,MHz. Using this method we are better able to account for the varying flux scale across the image, as can be seen by comparing the MWA flux densities to the expected flux density in the two test sources PMN\,J2004-5534 and PMN\,J2017-5747 (Fig.~\ref{im:corrected}, Table~\ref{tab:fluxresults}).

\begin{table*}
\setlength{\tabcolsep}{3pt}
 \caption{A comparison between the fitted flux density taken from archival data and the flux density in the MWA bands after applying the interpolation flux density correction method described in the text. The residual flux density errors range between 2 and 19\% with an average of 9\%.}
\begin{tabular}{l|cccc|cccc|cccc}
\hline
 Source & \multicolumn{4}{c|}{Expected Flux Density Jy\,beam$^{-1}$ } & \multicolumn{4}{c|}{Corrected MWA Flux Density Jy\,beam$^{-1}$} & \multicolumn{4}{c}{Residual Error (\%)} \\
 		Name & 120\,MHz & 149\,MHz & 180\,MHz & 226\,MHz & 120\,MHz & 149\,MHz & 180\,MHz & 226\,MHz & 120\,MHz & 149\,MHz & 180\,MHz & 226\,MHz  \\		
 	\hline
 	
PMNJ2017-5747 & 7.63 & 5.90 & 4.28 & 3.59 & 8.03 & 6.53 & 4.79 & 4.27 & 5.2 & 10.3 & 12.1 & 19.3 \\
PMNJ2004-5534 & 3.72 & 3.04 & 2.37 & 2.07 & 3.24 & 2.79 & 2.31 & 2.01 & 12.9 & 8.4 & 2.4 & 2.3 \\

 \hline
 \end{tabular}
 \label{tab:fluxresults}
 \end{table*}
 
This method is dependent on the number of sources available for interpolation and the accuracy of the flux density estimate for each source taken from the literature. Due to the low number of sources available, we are conservative in our flux density error estimate at each band. The residual flux density offset after applying this interpolation method ranges from 2--19\% with an average error of 9\% (Table~\ref{tab:fluxresults}). We use an average of the offsets for the two test sources PMNJ2017-5747 and PMNJ2004-5534 at each frequency for our flux density calibration errors. This gives a percentage flux density error of 9.1\%, 9.4\%, 4.9\% and 10.8\% at 120, 149, 180 and 226\,MHz, respectively. From the grid of correction values in Fig.~\ref{im:correction}, we find that the value of the flux density offset is approximately constant over the area of the NW relic of A3667 (0\farcm5). This method leads to slightly non-Gaussian background noise on large scales and an elevated noise level of 83.3, 50.0, 39.5 and 35.1\,mJy\,beam$^{-1}$ at 120, 149, 180, 226\,MHz, respectively. However, because this technique is multiplicative, the signal-to-noise ratio remains the same as for the uncorrected images.

\label{lastpage}

\end{document}